\documentstyle[epsfig,longtable]{aipproc_btev}

\def\Journal#1#2#3#4{{#1} {\bf #2}, #3 (#4)}

\def\NIM{\em Nucl. Instrum. Methods}

\def\PLB{{\em Phys. Lett.}  B}
\def\PRL{\em Phys. Rev. Lett.}
\def\PRD{{\em Phys. Rev.} D}

\def\be{\begin{equation}}
\def\ee{\end{equation}}
\def\bea{\begin{eqnarray}}
\def\eea{\end{eqnarray}}

\def\etal{{\it et al}}

\begin{document}

\title{{\normalsize\rightline{SUHEP-98/10}\rightline{Sept. 1998}
\rightline{ICHEP XXIX}}\vspace {1cm}  THE BTeV 
PROGRAM at 
Fermilab}

\tighten
\author{\large The BTeV Collaboration}
\normalsize
\author{A.~Kulyavtsev, M.~Procario, J.~Russ, and J.~You}
\address{Carnegie Mellon University, Pittsburgh PA 15213}
\author{J. Cumalat}
\address{University of Colorado at Boulder, Boulder CO 80309-0390}
\author{J.A.~Appel, C.N.~Brown, J.~Butler, H.~Cheung, D.~Christian, I.~Gaines,
P.~Garbincius, L.~Garren, N.~M.~Gelfand, G.~Jackson, Penny~Kasper,
Peter~Kasper, R.~Kutschke, S.~W.~Kwan, P.~Lebrun, P.~McBride, L.~Stutte,
and J.~Yarba}
\address{Fermi National Laboratory, Fermilab, P.O.Box 500, Batavia, IL 60510-0500}
\author{P.~Avery, and M.~Lohner}
\address{University of Florida at Gainesville, Gainesville, FL 32611-8440}
\author{R.A.~Burnstein, D.M.~Kaplan, L.M.~Lederman, H.A.~Rubin,
and C.~White}
\address{Illinois Institute of Technology, Chicago, IL 60616}
\author{M.~Selen, and J. Wiss}
\address{University of Illinois, Urbana-Champaign, IL 61801}
\author{R. Gardner, and J. Rust}
\address{Indiana University, Bloomington, IN 47405}
\author{D.~Menasce, L.~Moroni, D.~Pedrini, and S.~Sala}
\address{INFN and Dipartimento di Fisica dell'Universit`a and 
INFN - Milano, I-20133 Milan, Italy}
\author{G.~Boca, G.~Liguori, and P.~Torre }
\address{Dip. di Fisica Nucleare e Teorica and  INFN - Pavia, I-27100 Pavia, Italy }
\author{Y.~Kubota, and R~Poling}
 \address{University of Minnesota, Minneapolis, MN 55455}
\author{T.Y.~Chen}
\address{Nanjing University, Nanging 210008 P. R. China }
\author{V.~Papavassiliou}
\address{New Mexico State University, Las Cruces, NM 88003}
\author{P.~Maas}
\address{Northwestern University, Evanston, IL 60208-3112}
\author{K.~Honscheid, H.~Kagan, and A.~Wolf}
\address{Ohio State University, Columbus, OH 43210 }
\author{W.~Selove, and K.~Sterner}
\address{University of Pennsylvania, Philadelphia, PA 19104}
\author{A.~Lopez}
\address{University of Puerto Rico at Mayaguez, Mayaguez, PR 00681}
\author{M.~He}
\address{Shandong University, Jinan, Shandong 250100 P. R. China}
\author{S.~Shapiro}
\address{Stanford Linear Accelerator Center, Menlo Park, CA 94025}
\author{M.~Alam, and S.~Timm}
\address{State University of New York at Albany, Albany, NY 12222}
\author{M.~Artuso, M.~Goldberg, T.~Skwarnicki, S.~Stone, A.~Titov,
and J.C.~Wang}
\address{Syracuse University, Syracuse, NY 13244}
\author{T.~Handler}
\address{University of Tennessee, Knoxville, TN 37996-1200}
\author{A.~Napier}
\address{Tufts University, Medford, MA 02155}
\author{D.D.~Koetke}
\address{Valparaiso University,Valparaiso, IN 46383-6493}
\author{M.~Sheaff}
\address{University of Wisconsin, Madison WI 53706, and CINVESTAV, Mexico,
07360, D.F. , Mexico}
\author{X.Q.~Yu}
\address{University of Science and Technology of China, Hefei  230026 P. R. China}
\author{P.~Sheldon, and M.~Webster}
\address{Vanderbilt University, Nashville, TN 37235}
\author{J.~Slaughter}
\address{Yale University, New Haven, CT 06520-8120}
\author{S.~Menary}
\address{York University,~Toronto,~Ontario,~Canada~M3J~1P3}

\maketitle

\begin{abstract} {A  description is given of BTeV, a proposed program
at the Fermilab collider sited at the C0 intersection region. The main goals
are measurement of mixing, CP violation and rare decays in both the $b$ and
charm systems. The detector is a two-arm-forward spectrometer capable
of triggering on detached vertices and dileptons, and possessing excellent 
particle identification, electron, photon and muon detection.}
\end{abstract}

\section{Introduction}
BTeV is a Fermilab collider program whose main goals are to measure
mixing, CP violation and rare decays in the $b$ and $c$ systems. Using the new
Main injector, now under construction, the collider will produce on the order
of $4\times10^{11}$ $b$ hadrons in $10^7$ sec. of running. This compares 
favorably
with $e^+e^-$ colliders operating at the $\Upsilon$(4S) resonance. These
machines, at their design luminosities of $3\times 10^{33}$cm$^{-2}$s$^{-1}$
will produce $6\times 10^7$ $B$ mesons in $10^7$ seconds \cite{bfacs}. 

\section{Importance of Heavy Quark Decays}
The physical point-like states of nature that have both strong and electroweak
interactions, the quarks, are mixtures of base states described by the
Cabibbo-Kobayashi-Maskawa matrix:\cite{ckm}
\begin{equation}
\left(\begin{array}{c}d'\\s'\\b'\\\end{array} \right) =
\left(\begin{array}{ccc} 
V_{ud} &  V_{us} & V_{ub} \\
V_{cd} &  V_{cs} & V_{cb} \\
V_{td} &  V_{ts} & V_{tb}  \end{array}\right)
\left(\begin{array}{c}d\\s\\b\\\end{array}\right)
\end{equation}
The unprimed states are the mass eigenstates, while the primed states denote
the weak eigenstates. A similar matrix describing neutrino mixing is possible
if the neutrinos are not massless.

There are nine complex CKM elements. These 18 
numbers can be reduced to four independent quantities by applying unitarity 
constraints and using the fact that the phases of the quark wave functions are 
arbitrary. 
These four remaining numbers are  fundamental constants of nature that 
need to be determined experimentally, like any other
fundamental constant such as $\alpha$ or $G$. In the Wolfenstein 
approximation the matrix is written as\cite{wolf}
\begin{equation}
V_{CKM} = \left(\begin{array}{ccc} 
1-\lambda^2/2 &  \lambda & A\lambda^3(\rho-i\eta) \\
-\lambda &  1-\lambda^2/2 & A\lambda^2 \\
A\lambda^3(1-\rho-i\eta) &  -A\lambda^2& 1  
\end{array}\right).
\end{equation}
The constants $\lambda$ and $A$ have been measured \cite{virgin}.

The phase $\eta$ allows for CP violation. 
CP violation thus far has been seen only in the neutral kaon 
system. If we can find CP violation in the $B$ system we could see
if the CKM  model works or perhaps go beyond the model. Speculation has it that
CP violation  is  responsible for the baryon-antibaryon asymmetry in our
section of the Universe. If  so,  to understand the mechanism of CP violation
is critical in our conjectures of why  we  exist \cite{langacker}.

\begin{figure}[hbtp]
\vspace{-6mm}
\centerline{\epsfig{figure=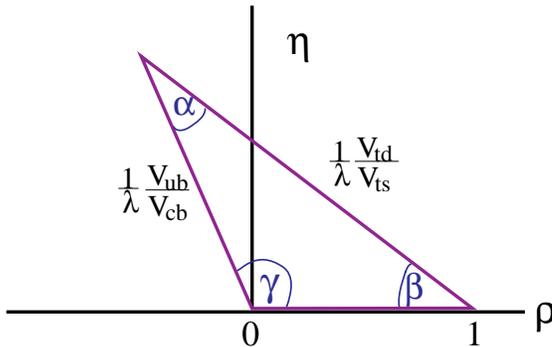,height=2.in}}
\vspace{-4mm}
\caption{The unitarity triangle shown in the $\rho-\eta$ plane. The
left side is determined by measurements of $b\to u/b\to c$ and the right side 
can be
determined using mixing measurements in the $B_s$  and $B_d$ systems. The 
angles can be found by making measurements of CP violating asymmetries in
hadronic $B$ decays.
\label{ut_tri}}
\end{figure}

Unitarity of the CKM matrix leads to the constraint triangle
shown in Fig.~\ref{ut_tri}. The left side can be measured using charmless
semileptonic $b$ decays, while the right side can be measured by using the ratio
of $B_s$ to $B_d$ mixing. The angles can be found by measuring CP violating
asymmetries in hadronic $B$ decays.

\begin{figure}[htb]
\vspace{-.04cm}
\centerline{\epsfig{figure=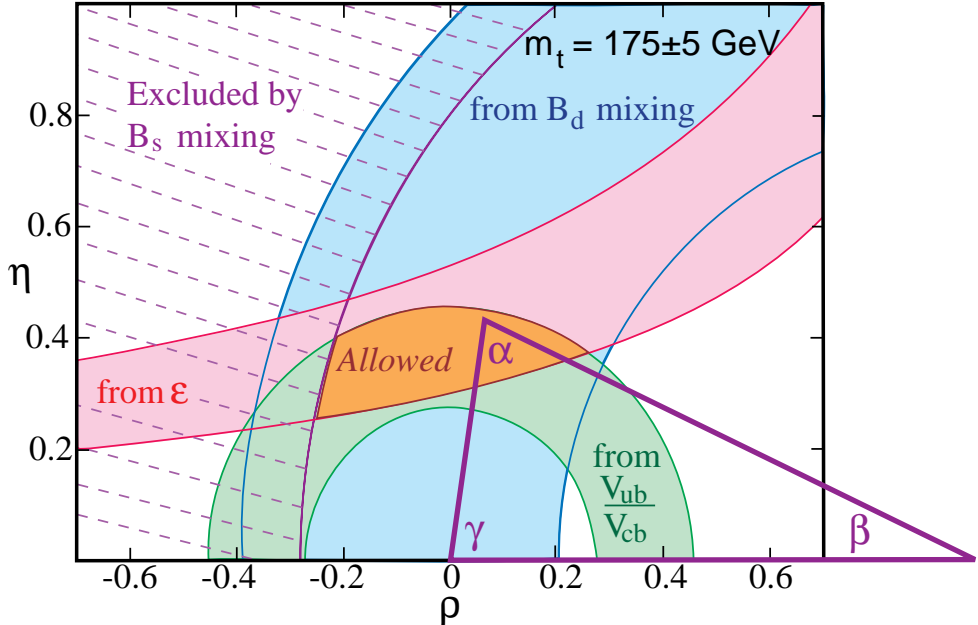,height=3.3in}}
\vspace{-.1cm}
\caption{\label{ckm_fig}The regions in $\rho-\eta$ space (shaded) consistent
with measurements of CP violation in $K_L^o$ decay ($\epsilon$), $V_{ub}/V_{cb}$
in semileptonic $B$ decay, $B_d^o$ mixing, and the excluded region from
limits on $B_s^o$ mixing. The allowed region is defined by the overlap of
the 3 permitted areas, and is where the apex of the  CKM triangle  sits. The
bands represent $\pm 1\sigma$ errors. The error on the $B_d$ mixing band is
dominated by the parameter $f_B$. Here the range is taken as 
$240> f_B > 160$ MeV.}
\end{figure}

The current status of constraints on $\rho$ and $\eta$ is shown in 
Fig.~\ref{ckm_fig}. One constraint on $\rho$ and $\eta$ is given by the $K_L^o$ CP
violation  measurement ($\epsilon$) \cite{buras}, where the largest error arises
from theoretical uncertainty. Other constraints come from current measurements
on $V_{ub}/V_{cb}$, and $B_d$  mixing \cite{virgin}. The widths of both of these
bands are dominated by theoretical errors. Note that the errors used are
$\pm 1\sigma$. This shows that the data are consistent with the standard model
but do not pin down $\rho$ and $\eta$.

It is crucial to check if measurements of the sides and angles are consistent,
i.e., whether or not they actually form a triangle. The standard model is
incomplete. It has many parameters including the four CKM numbers, six quark
masses, gauge boson masses and coupling constants. Perhaps measurements of the
angles and sides of the unitarity triangle will bring us beyond the standard
model and show us how these parameters are related, or what is missing. 

\section{The Main Physics Goals of BTeV}
\subsection{Physics Goals For B's}
Here we briefly list the main physics goals of BTeV for studies of the $b$ 
quark.

$\bullet$ Precision measurements of $B_s$ mixing, both the time evolution,
$x_s$, and the lifetime difference, $\Delta\Gamma$, between the positive CP and
negative CP final states.

$\bullet$ Measurement of the ``CP violating" angles $\alpha$ and $\gamma$. We
will use $B^o\to \pi^+\pi^-$ for $\alpha$ \cite{alphahard} and measure $\gamma$
using several different methods including measuring the time dependent asymmetry 
in $B_s^o\to D_s^{\pm}K^{\mp}$, and measuring the decay rates $B^+\to D^oK^+$  
and $B^-\to
\overline{D}^oK^+$, where the $D^o$ can decay directly or via a doubly Cabibbo
suppressed decay mode \cite{gronau,sad}.

$\bullet$ Search for rare final states such as $K\mu^+\mu^-$ and $\pi\mu^+\mu^-$
which could result from new high mass particles coupling to $b$ quarks.

$\bullet$ We assume that the CP violating angle $\beta$ will have already been
measured by using $B^o\to \psi K_s$, but we will be able to significantly
reduce the error. 

\subsection{The Main Physics Goals for charm}
According to the standard model, charm mixing and CP violating effects should
be ``small." Thus charm provides an excellent place for non-standard model
effects to appear. Specific goals are listed below.

$\bullet$ Search for mixing in $D^o$ decay, by looking for both the rate of 
wrong sign decay, $r_D$, and the width difference between positive CP and
negative CP eigenstate decays, $\Delta\Gamma$. The current upper limit on $r_D$
is $3.7\times 10^{-3}$, while the standard model expectation is
$r_D<10^{-7}$ \cite{dmix}.

$\bullet$ Search for CP violation in $D^o$. Here we have the advantage over $b$
decays that there is a large $D^{*+}$ signal which tags the initial flavor of
the $D^o$ through the decay $D^{*+}\to \pi^+ D^o$. Similarly $D^{*-}$ decays
tag the flavor of initial $\overline{D}^o.$ The current experimental upper 
limits on CP violating asymmetries are on the order of 10\%, while the standard
model prediction is about 0.1\% \cite{dcp}.

$\bullet$ Search for direct CP violation in charm using $D^+$ and $D_s^+$ 
decays.

$\bullet$ Search for rare decays of charm, which if found would signal new
physics.

\subsection{Other $b$ and charm Physics Goals}
There are many other physics topics that can be addressed by BTeV. A short list
is given here.

$\bullet$ Measurement of the $b\overline{b}$ production cross section and
correlations between the $b$ and the $\overline{b}$ in the forward direction.

$\bullet$ Measurement of the $B_c$ production cross section and decays.

$\bullet$ The spectroscopy of $b$ baryons.

$\bullet$ Precision measurement of $V_{cb}$ using the usual mesonic decay
modes and the baryonic decay mode
$\Lambda_b\to \Lambda_c\ell^-\bar{\nu}$ to check the form factor shape
predictions.

$\bullet$ Precision measurement of $V_{ub}/V_{cb}$ using the usual mesonic decay 
modes.

$\bullet$ Measurement of the $c\overline{c}$ production cross section and
correlations between the $c$ and the $\overline{c}$ in the forward direction.

$\bullet$ Precision measurement of $V_{cd}$ and the form factors in the
decays $D\to\pi\ell^+\nu$ and $D\to\rho\ell^+\nu$.

$\bullet$ Precision measurement of $V_{cs}$ and the form factors in the
decay $D\to K^*\ell^+\nu$.

\section{Characteristics of Hadronic $b$ Production}

It is often customary to characterize heavy quark production in hadron
collisions with the two variables $p_t$ and $\eta$. The latter variable was
invented by those who studied high energy cosmic rays and is assigned the
value 
\begin{equation}
\eta = -ln\left(\tan\left({\theta/2}\right)\right),
\end{equation}
where $\theta$ is the angle of the particle with respect to the beam direction.

\begin{figure}[htb]
\vspace{-1.3cm}
\centerline{\epsfig{figure=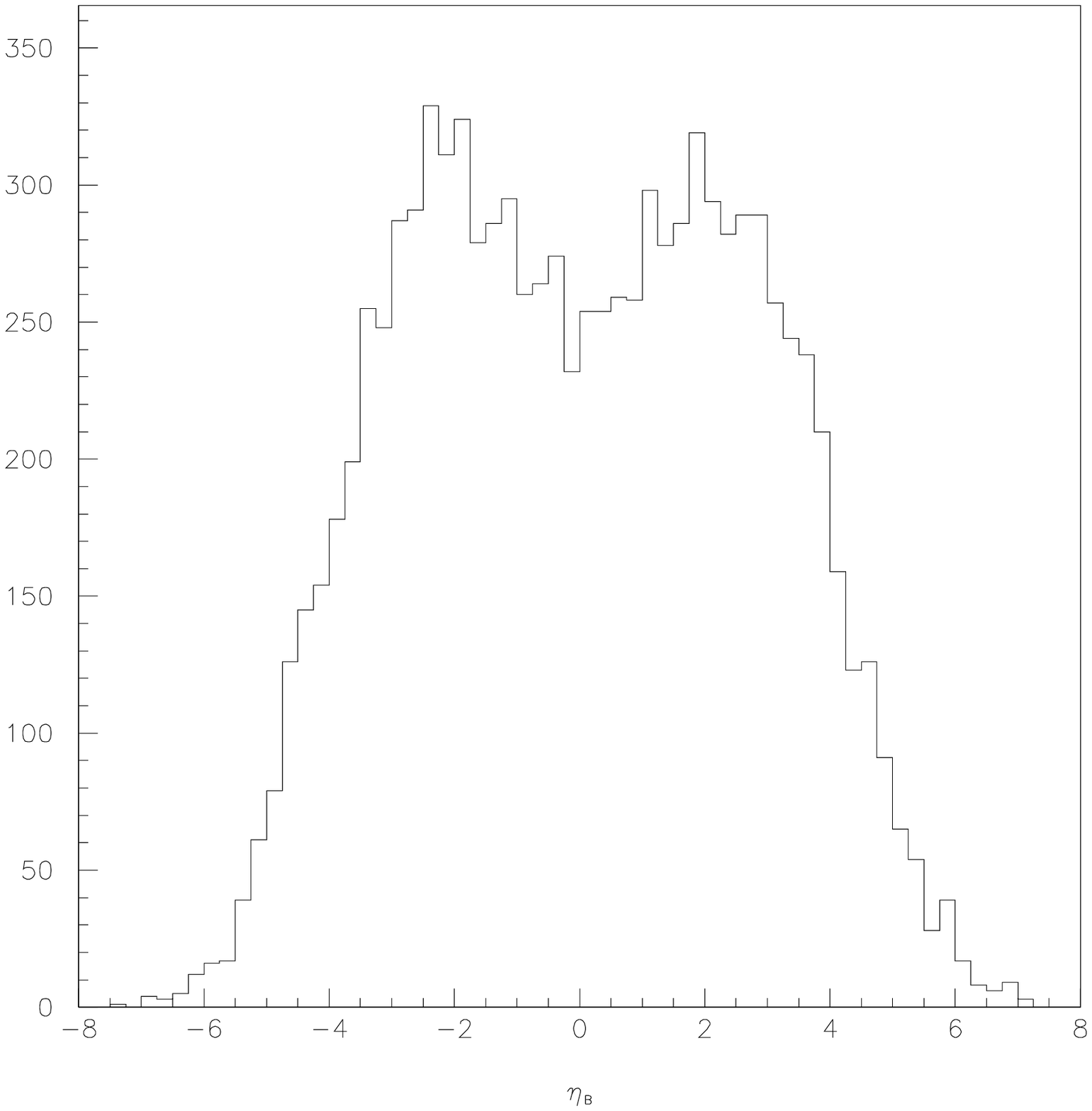,width=2.5in}
\epsfig{figure=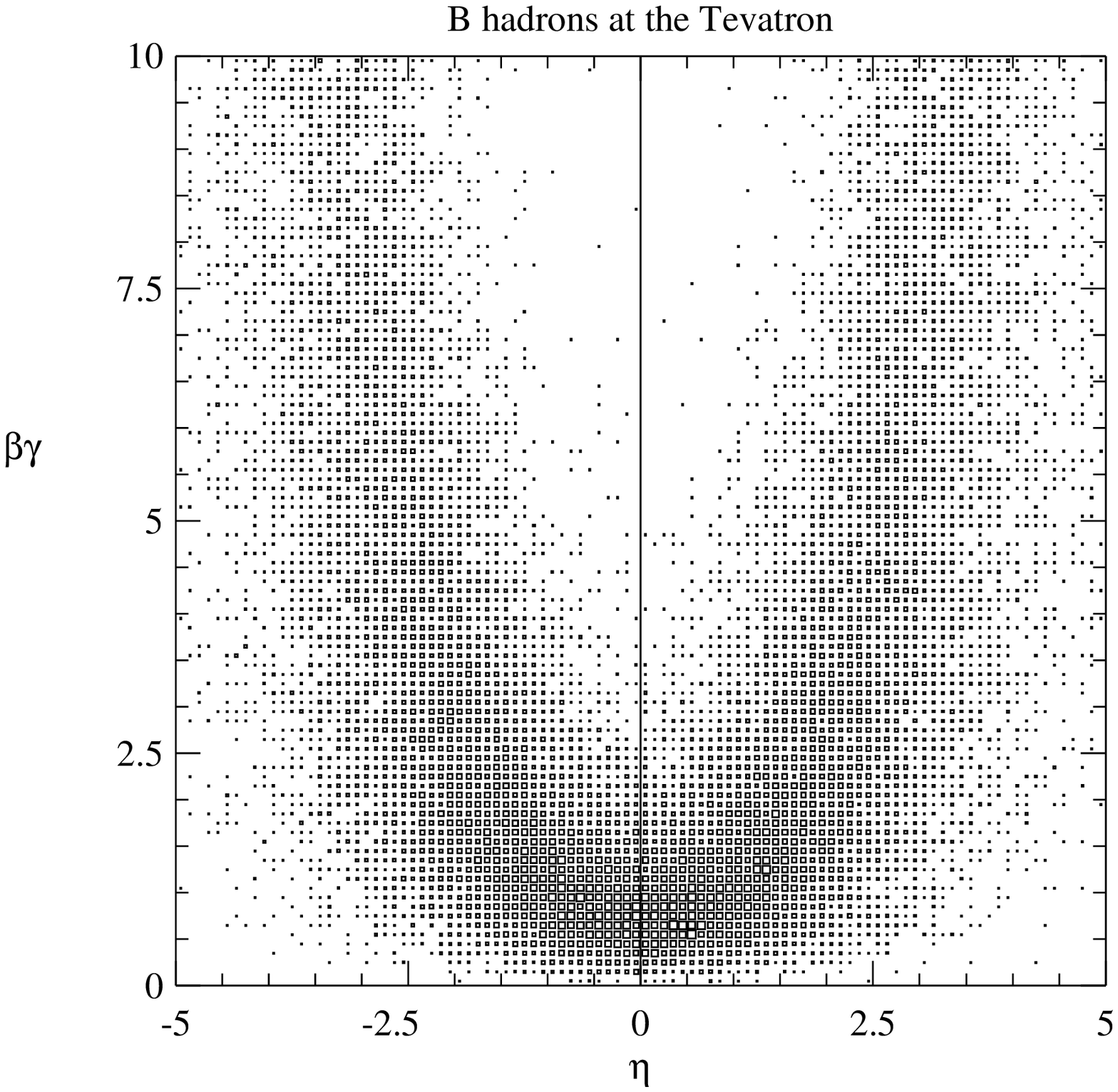,width=2.6in}}
\vspace{-.1cm}
\caption{\label{n_vs_eta}  The $B$ yield versus $\eta$ (left). 
$\beta\gamma$ of the  $B$  versus $\eta$ (right). Both plots are for the
Tevatron.}
\end{figure}

According to QCD-based calculations of $b$-quark production, the $b$'s are
produced ``uniformly" in $\eta$ and have a truncated transverse momentum,
$p_t$, spectrum, characterized by a mean value approximately equal to the $B$
mass \cite{artuso}. The distribution in $\eta$ is shown in Fig.~\ref{n_vs_eta}.

There is a strong correlation between the $B$ momentum and $\eta$. Shown also in
Fig.~\ref{n_vs_eta} is the $\beta\gamma$ of the $B$ hadron versus $\eta$.
It can clearly be seen that near $\eta$ of zero, $\beta\gamma\approx 1/2$, while
at larger values of $|\eta |$, $\beta\gamma$ can easily reach values of ~6.
This is important because the observed decay length increases with $\beta\gamma$
and furthermore the absolute momenta of the decay products are larger allowing
for a suppression of the multiple scattering error.

\begin{figure}[htb]
\vspace{-.2cm}
\centerline{\epsfig{figure=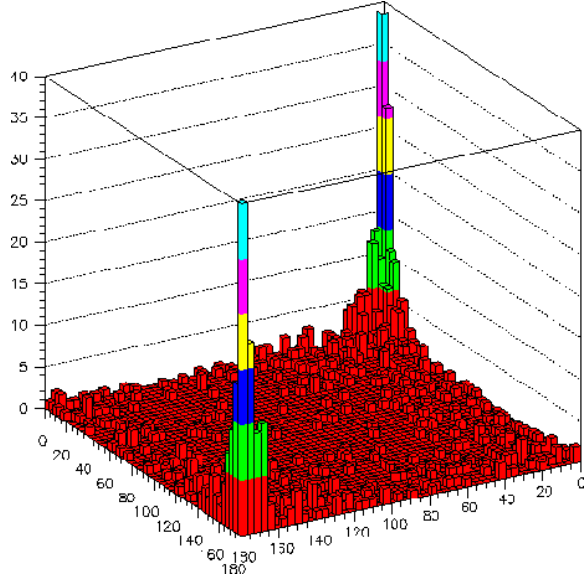,height=3.0in}}
\caption{\label{bbar}The production angle (in degrees) for the hadron
containing a $b$ quark plotted versus the production angle for the hadron
containing a $\bar{b}$ quark.}
\end{figure}

The ``flat" $\eta$ distribution hides an important correlation of $b\bar{b}$
production at hadron colliders. In Fig.~\ref{bbar} the production angle of
the hadron containing a $b$ quark is plotted versus the production angle of
the hadron containing a $\bar{b}$ quark according to the Pythia generator.
There is a very strong correlation in the forward direction (the direction of
the $p$ beam at 0$^{\circ}$-0$^{\circ}$), where both $B$ and $\overline{B}$
hadrons are going in the same direction. The same strong correlation is present
in the $\overline{p}$ direction. This correlation is not present in the
central region (near 90$^{\circ}$). By instrumenting a relatively small region of
angular phase space, a large number of $b\bar{b}$ pairs can be detected.
Furthermore the $B$'s populating the two ``forward" regions have large values
of $\beta\gamma$.

Charm production is similar to $b$ production but more copius. Current
theoretical estimates are that charm is 1-2\% of the total $p\bar{p}$
cross section. 

Table ~\ref{tab:b_c} gives the relevant Tevatron parameters.
We expect to eventually run at a luminosity of $2\times 10^{32}$cm$^{-2}$s$^{-1}$. 
A machine design that holds the luminosity constant at this value, called 
``luminosity leveling," has been developed. We plan to adopt this design.
\begin{table}
\caption{The Tevatron as a $b$ and $c$ source for C0 in Run II.\label{tab:b_c}}
\vspace{0.4cm}
\begin{center}
\begin{tabular}{lc}  
Luminosity  & $2\times 10^{32}$ cm$^{-2}$s$^{-1}$\\
$b$ cross section & 100 $\mu$b \\ 
\# of $b$'s per 10$^7$ sec  & $4\times 10^{11}$\\
$b$ fraction & 0.2\% \\
$c$ cross section & $>500~\mu$b \\
Bunch spacing & 132 ns \\
Luminous region length & $\sigma_z$ = 30 cm\\
Luminous region length & $\sigma_x$ $\sigma_y$ = $\approx 50$ $\mu$m\\
Interactions/crossing & $<2>$
 \\ \hline
\end{tabular}
\end{center}
\end{table}

\section{The Experimental Technique: A Forward Two-arm Spectrometer}

\begin{figure}[htb] 
\centerline{\epsfig{figure=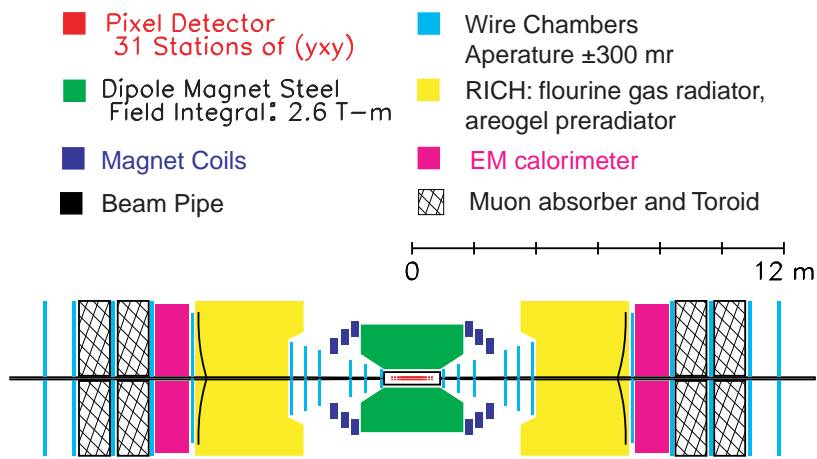,height=3.2in}}
\caption{\label{btev_det_doc}Sketch of the BTeV spectrometer.}
\end{figure}

A sketch of the apparatus is shown in Fig.~\ref{btev_det_doc}. The two-arm
spectrometer fits in the expanded C0 interaction region, which is being
excavated. The magnet that we will use, called SM3, exists at
Fermilab. The other important parts of the experiment include the vertex
detector, the RICH detectors, the EM calorimeters and the muon system.  

\begin{figure}[hbt] 
\vspace{1 mm}
\centerline{\epsfig{figure=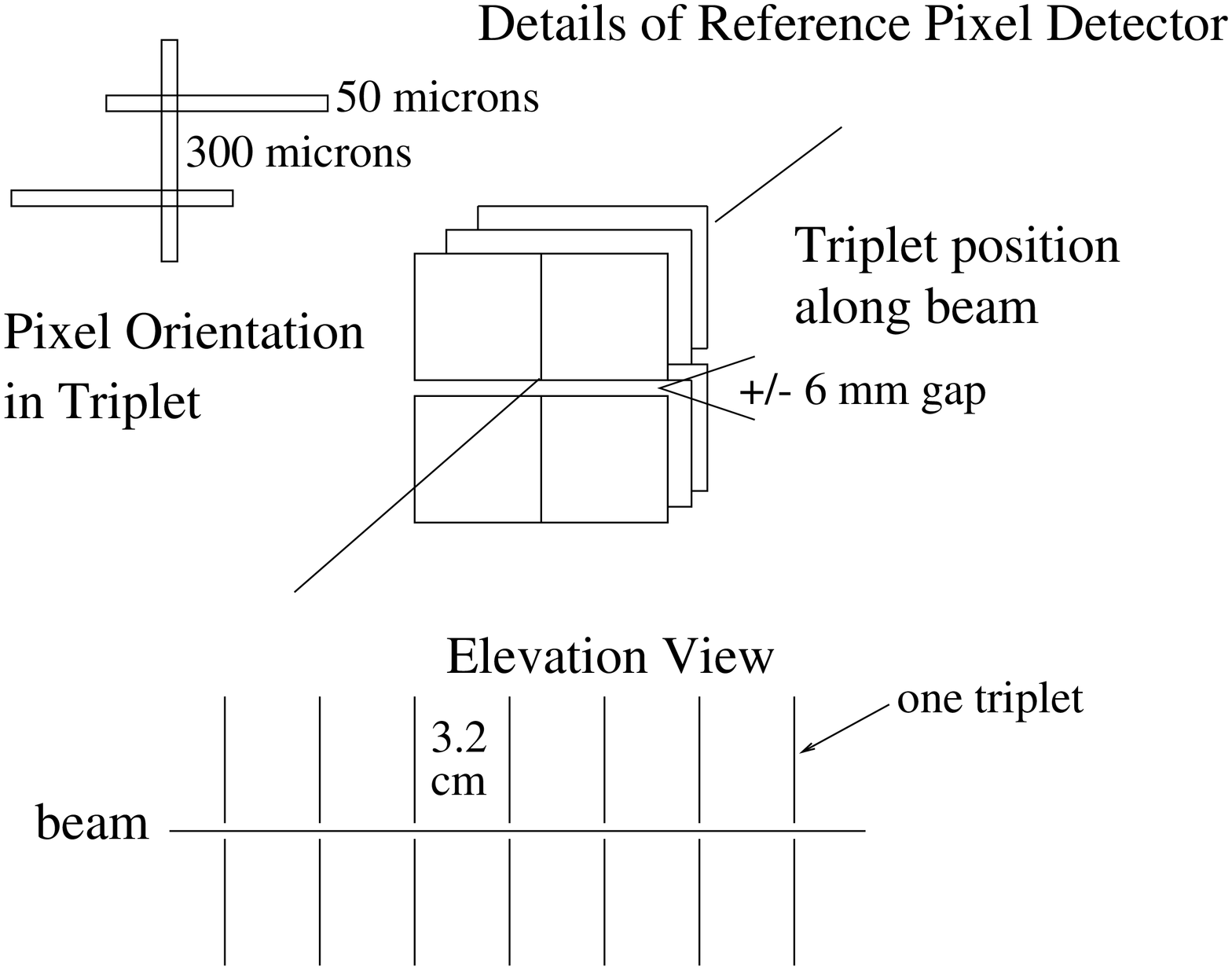,height=3.1in,angle=90}}
\caption{\label{vertex_50}Layout of the BTeV  pixel detector. There are 
31 stations of triplets with the narrow pixel dimension in the bend plane. The 
most recent version replaces the 12 mm gap between detector planes with a 12 mm
$\times$ 12 mm square hole centered on the beam.}
\vspace{1 mm} 
\end{figure}

The angle subtended is approximately $\pm$300 mr in both plan and
elevation views. The vertex detector is a multi-plane pixel device that sits
inside the beam pipe. The baseline design has 31 stations with triplets in each 
station. The detector is sketched in Fig.~\ref{vertex_50}. Our new baseline 
detector has a square hole, 12 mm $\times$ 12 mm around the beam, instead of a 12 
mm gap between top and bottom halves. (Some of our simulations have been done 
with the detector with the gap, called ``EOI," and some have been done with the 
``square hole.")
The triggering concept is to pipeline the data and to 
detect detached $b$ or $c$ 
vertices in the first trigger level. The vertex
detector is put in the magnetic field in order to insure that the tracks
considered for vertex based triggers do not have large multiple scattering
because they are low momentum.

The RICH detector\cite{tomasz} has a gas radiator, either C$_4$F$_{10}$ or
C$_5$F$_{12}$, and mirrors that focus the Cherenkov light onto photodectors
situated outside of the fiducial volume of the detector. This system will
provide $K/\pi$ separation in the momentum range between 3 and 70 GeV/c. To
resolve protons from kaons below the kaon threshold of 9 GeV/c, a thin aerogel
radiator may be placed in front of the gas volume. The same photon detector
would be utilized.

The muon detector consists of position-measuring chambers placed around and
between an iron slab followed by another slab used as a magnetized toroid. This
system is used both to trigger on final states with dimuons and to identify
muons in the final analysis.

\section{Simulations}

\subsection{Introduction}

We have developed several fast simulation packages to verify the basic BTeV
concepts and aid in the final design. The trigger simulations, discussed below,
are done with full pattern recognition. The input consists only of hits which
are smeared by their resolution. To simulate backgrounds in the final physics
analysis, we use a fast simulation which simulates track resolutions but not
the pattern recognition. This is done because we have to simulate backgrounds
in processes with branching ratios in the 10$^{-5}$-10$^{-6}$ range and we
cannot afford the computer time. The key program in our system is
MCFast \cite{MCFast}. Charged tracks are generated and traced through different
material volumes including detector resolution, multiple scattering and
efficiency. This allows us to measure acceptances and resolutions in a fast
reliable manner.

\subsection{Trigger Simulations}

We simulate the trigger using the baseline pixel detector shown in
Fig.~\ref{vertex_50}. The triplet 
stations each provide a three-dimensional space point as well as a track 
direction mini-vector. This 
is useful for fast pattern recognition. The trigger simulations are carried out 
by doing the complete 
pattern recognition from the hits left in the detector by tracks and converted 
photons.

\begin{figure}[hbt]
\vspace{-.3in}
\centerline{\epsfig{file=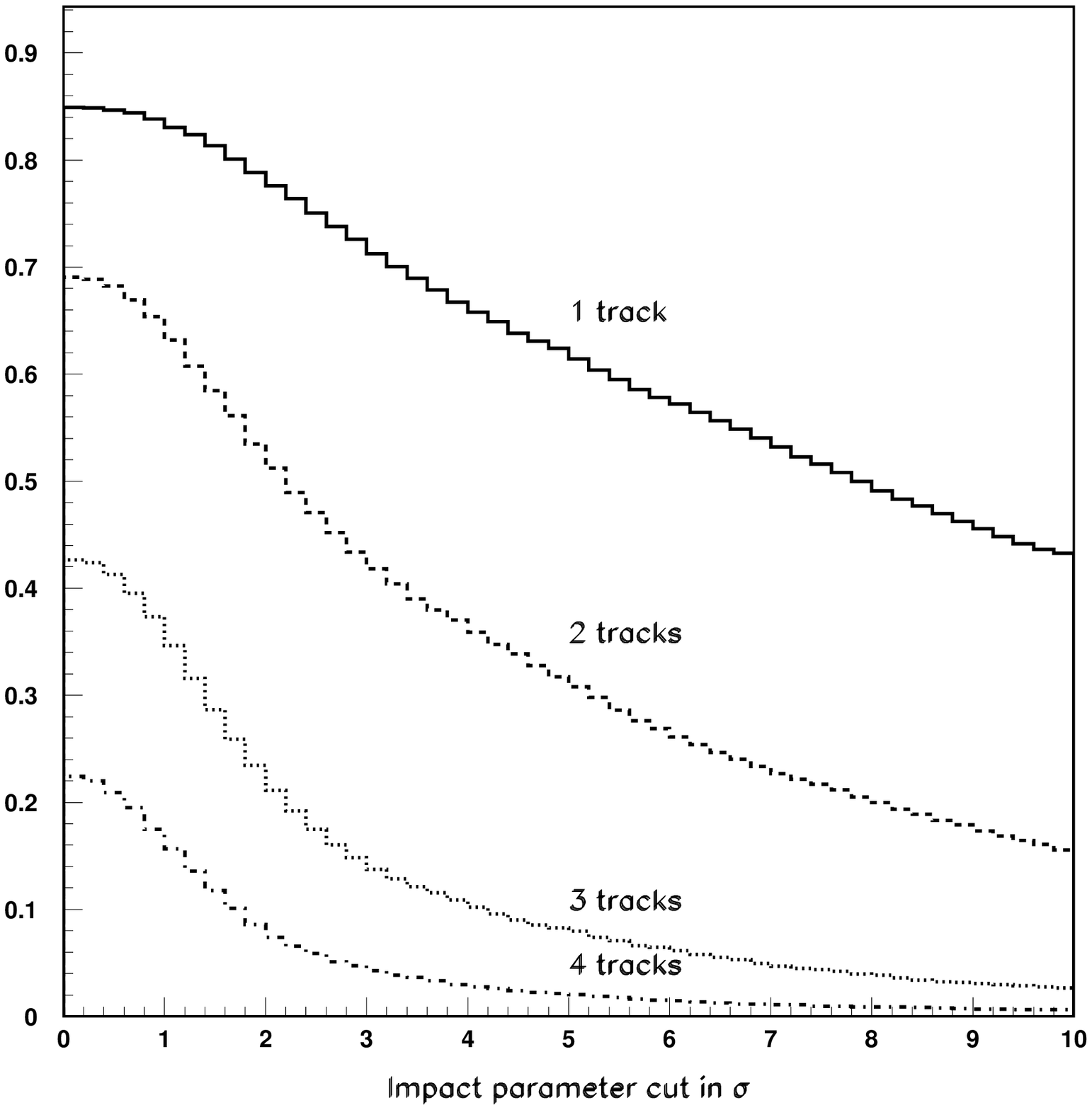,width=2.8in}
\epsfig{file=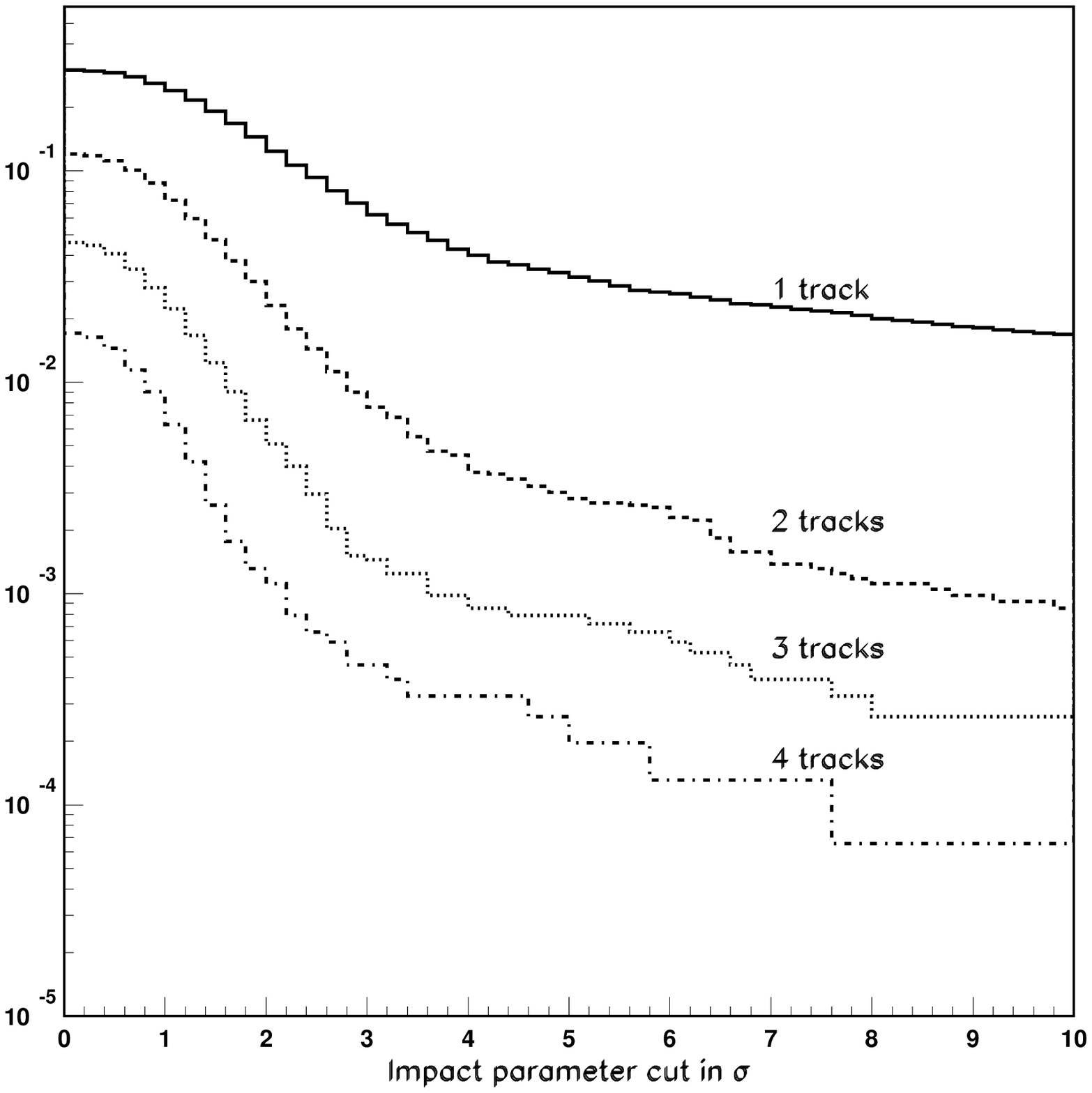,width=2.8in}}
\vspace{-.1cm}
\caption{\label{trig1} (left) Trigger efficiency for $B^o\to\pi^+\pi^-$ for 
pion tracks 
in the spectrometer. (right) Trigger efficiency for light quark events.
 The ordinate gives the choice of cut value
on the impact parameter in terms of number of standard deviations ($\sigma$) 
of the track from the primary vertex. The curves show the effect of requiring
 different numbers of tracks. }

\end{figure}

Our baseline trigger algorithm works by first determining the main event vertex
and then finding how many tracks miss this vertex by $n\sigma$, where $\sigma$
refers to the impact parameter divided by its error. Furthermore, a requirement
is then placed on the track momentum in the bend plane, $p_y$, as determined on
line. The preliminary results of simulating this algorithm are shown in
Fig.~\ref{trig1} (right) for a cut $p_y~>~0.5$ GeV/c \cite{Selove}. The choice 
of the number of tracks and the impact parameter requirement must eventually 
be optimized, but what is shown here is the efficiency for accepting light
quark events ($u$, $d$, and $s$) for various choices on the number of tracks
(curves) and the size of their required impact parameter divided by the  error
in impact parameter. The efficiency for accepting  $B^o\to\pi^+\pi^-$ is shown
in the left side. Here the efficiency is given after requiring that both tracks
are in the  spectrometer and accepted for further analysis. For a ``typical"
$n\sigma$ cut of 3 and track requirement of 2, the $\pi^+\pi^-$ trigger
efficiency is about 45\%, while  the light quark background  has an efficiency
of about 0.8\%. Note, that we do not consider $c$ to be a background in this
experiment. For a ``typical" charm reaction the same trigger gives substantially
less than 1\% efficiency on charm, while the efficiency for two-body charm
decays is approximately 1\%.

\begin{figure}[hbt]
\vspace{-.3cm}
\begin{center}
\epsfig{figure=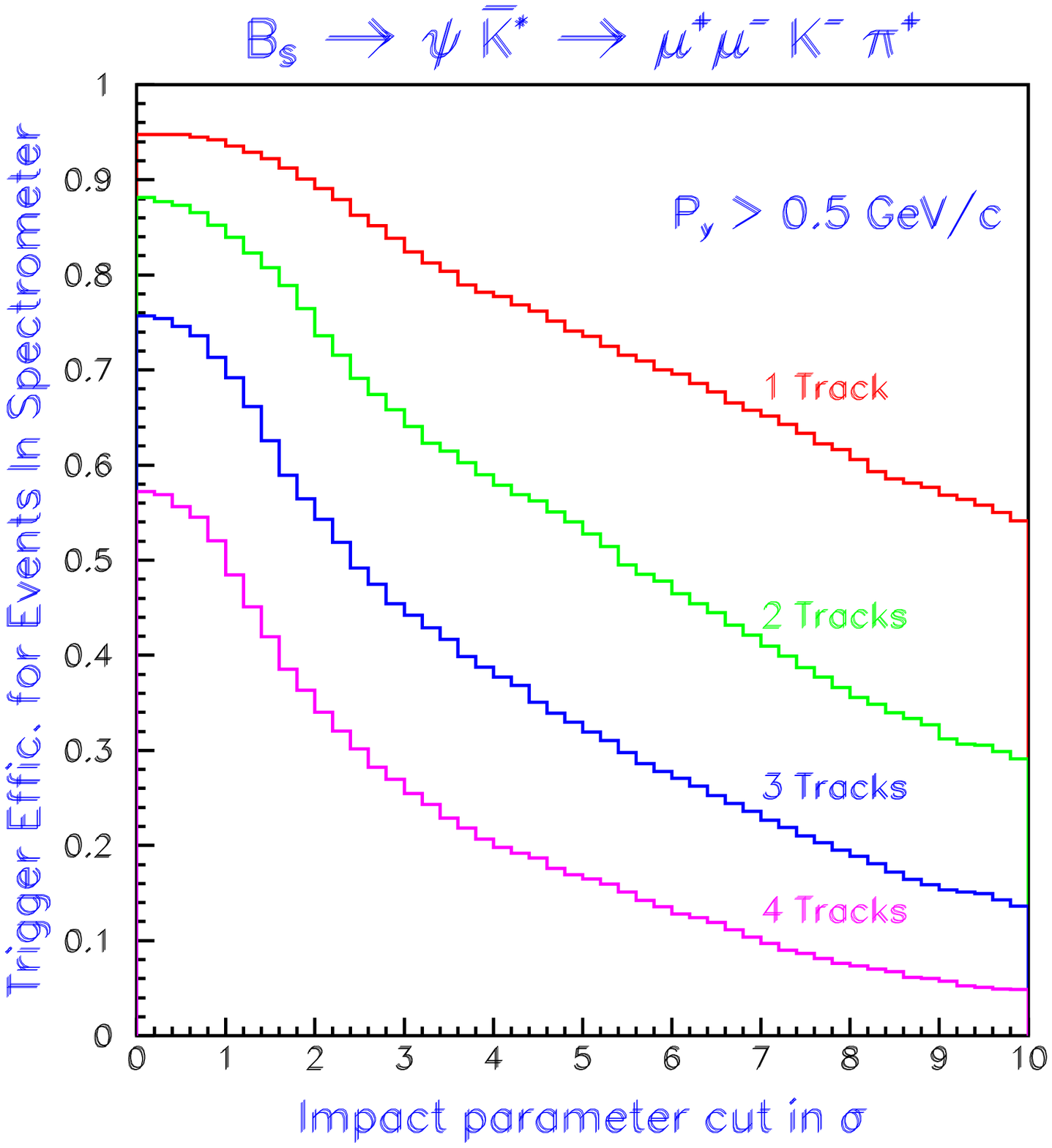,width=2.6in}
\epsfig{figure=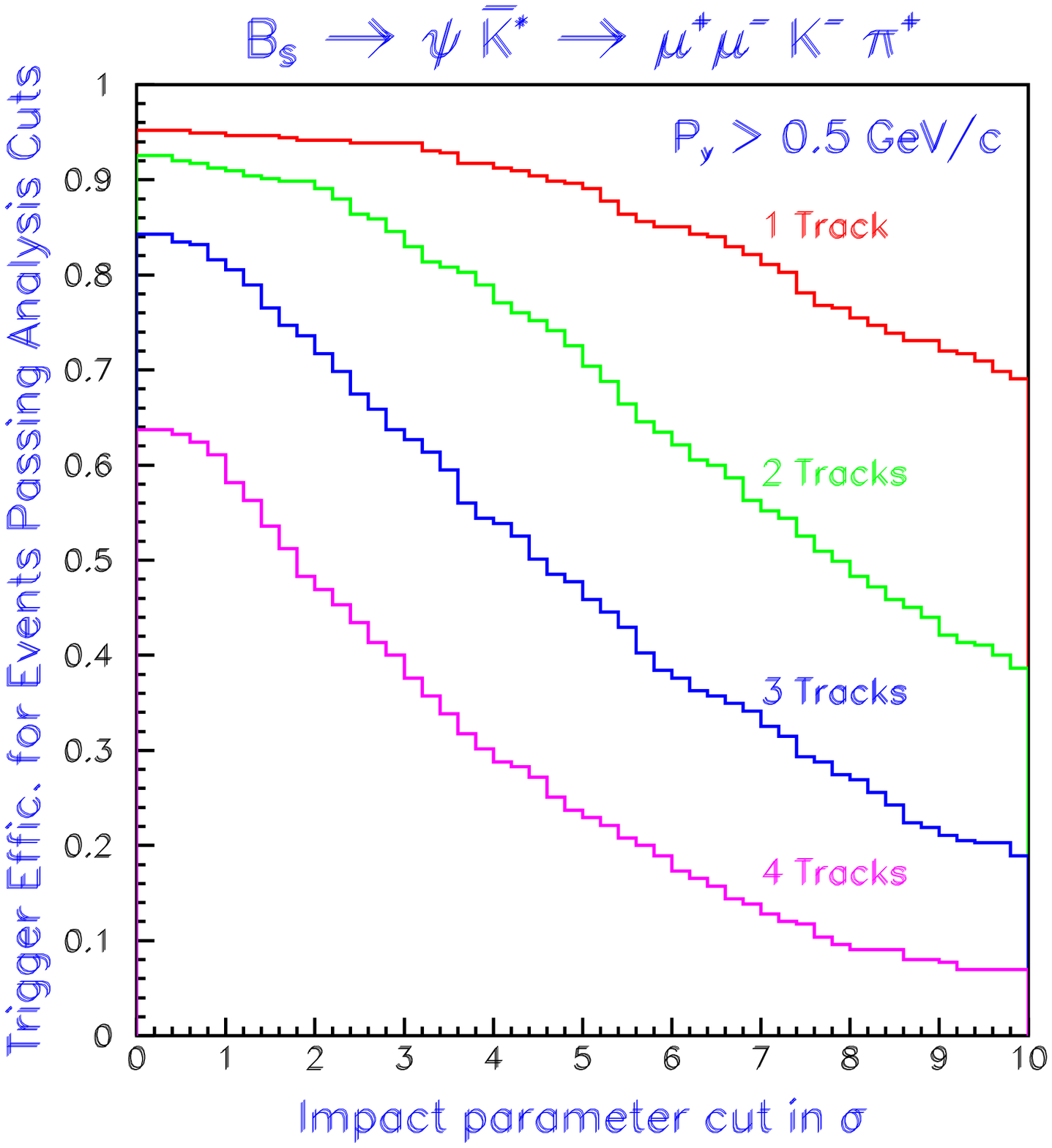,width=2.6in}
\end{center}
\caption{\label{psiks_trig}Trigger efficiency of $B_s\to\psi
K^{*o}$, $\psi\to\mu^+\mu^-$, $K^{*o}\to K^-\pi^+$ for tracks in the geometric
Acceptance (left) and after all analysis cuts (right).
The abscissa gives the value of the impact parameter in terms of number of 
standard deviations 
($\sigma$) 
of the track from the primary vertex. The curves show the effect of requiring
 different numbers of tracks. All tracks are required to have at least
 0.5 GeV/c momentum in the bend plane.}
\end{figure}

The trigger efficiency on states of interest is correlated with the analysis
criteria used to reject background. These criteria generally are focussed upon
insuring that the $B$ decay track candidates come from a detached vertex, that
the momentum vector from the $B$ point back to the primary interaction vertex,
and that there are no other tracks consistent with the $B$ vertex. When the
analysis criteria are applied first and the trigger efficiency evaluated after,
the trigger efficiency defined in this manner is larger. In
Fig.~\ref{psiks_trig} we show the efficiency to trigger on $B_s\to\psi K^{*o}$,
$\psi\to\mu^+\mu^-$, $K^{*o}\to K^-\pi^+$ using the tracking trigger only for
events with the four tracks in the geometric acceptance, and the efficiency
evaluated after all the analysis cuts have been applied. Here the trigger
efficiencies for $3\sigma$ and 2 tracks are 67\% for events with all 4 tracks
in the geometrical acceptance and 84\% on events after all the analysis cuts
have been applied.

\begin{figure}[hbt]
\begin{center}
\epsfig{figure=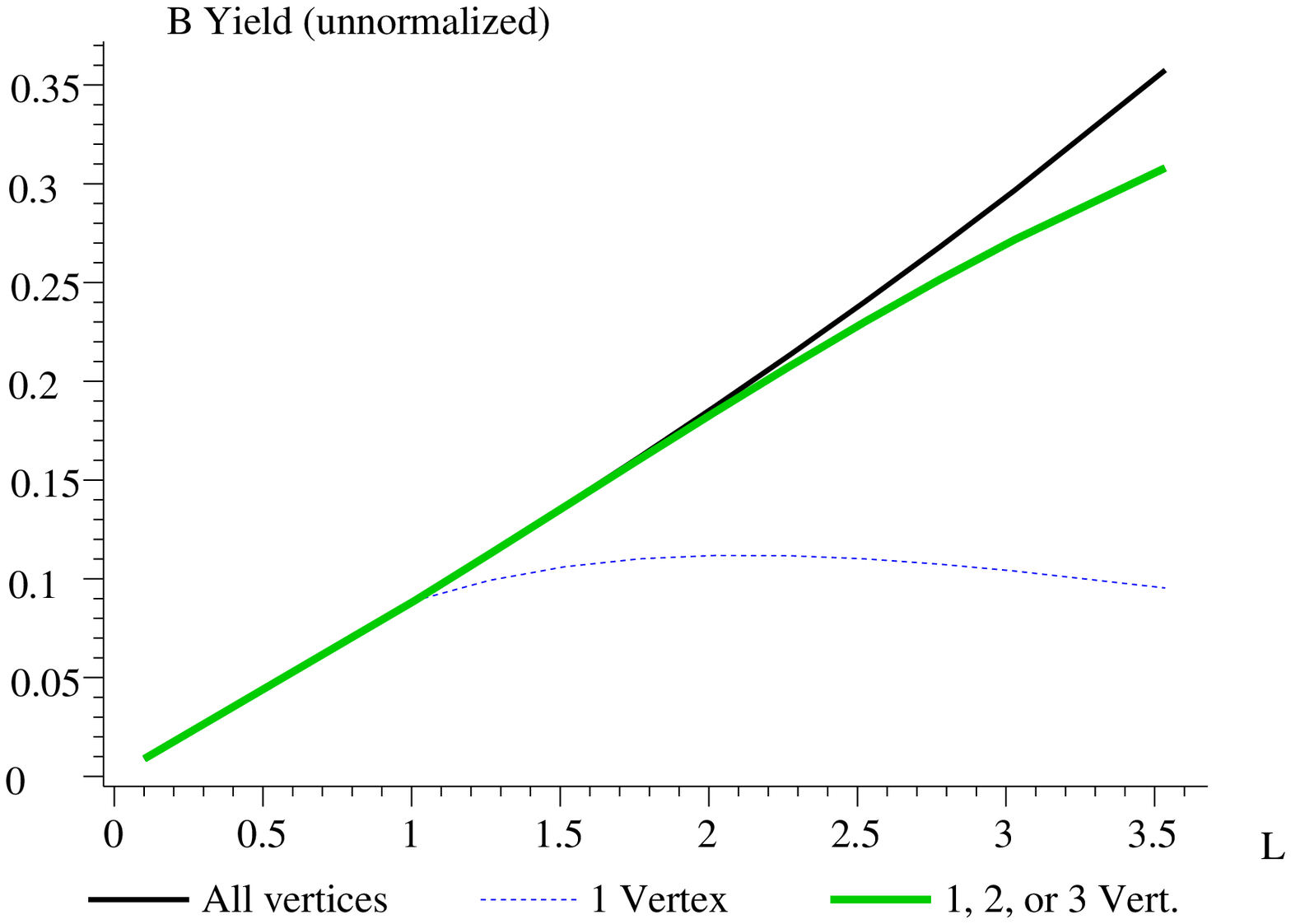,width=2.8in}
\epsfig{figure=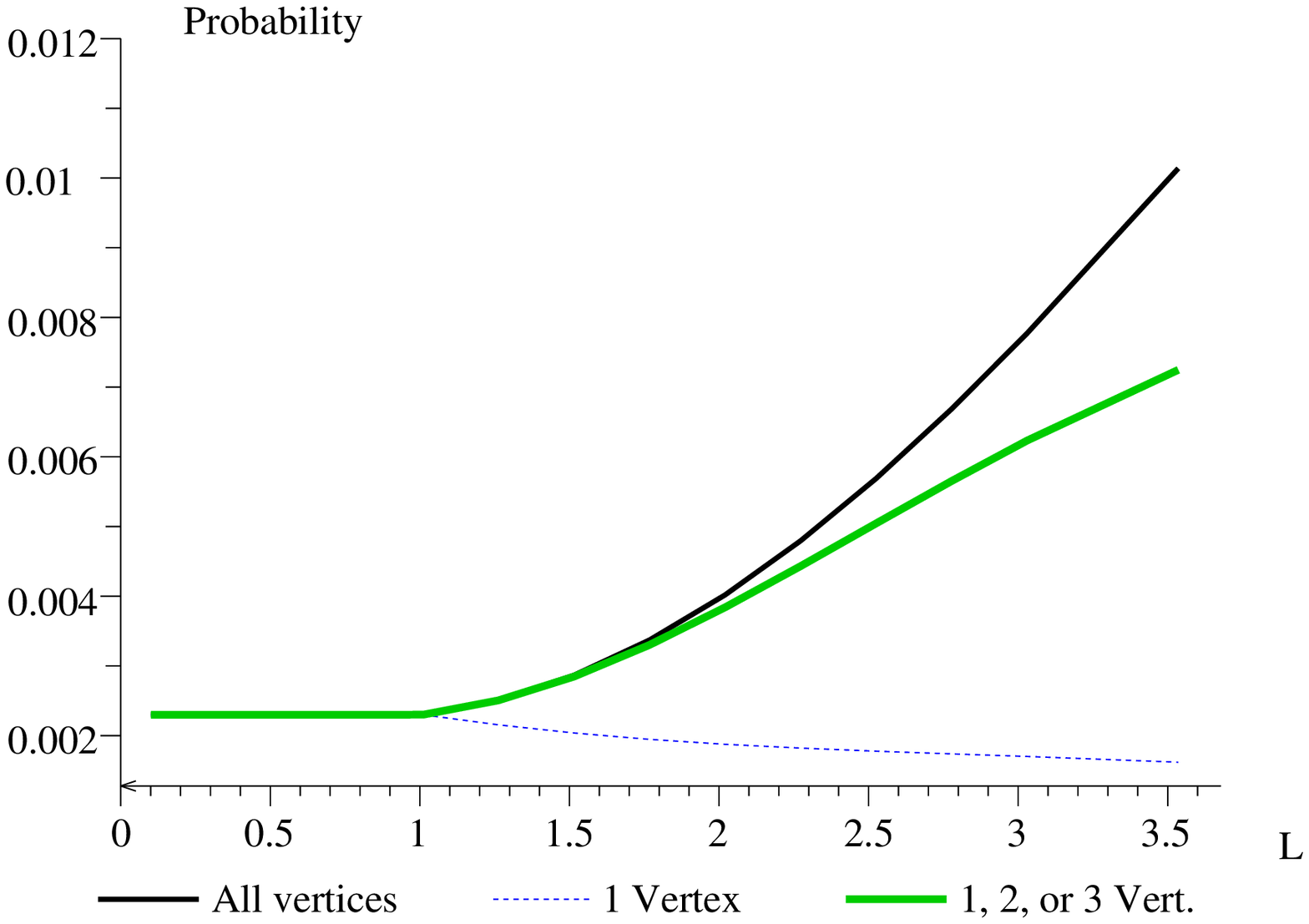,width=2.8in}
\end{center}
\caption{\label{lebrun_trig}Luminosity dependent trigger efficiencies for
$B^o\to\pi^+\pi^-$ (left), and light quark events (right). The abscissa is
in units of 10$^{32}$cm$^{-2}$s$^{-1}$. }
\end{figure}

At the BTeV design luminosity of 2$\times$10$^{32}$cm$^{-2}$s$^{-1}$ there is
an average of two interactions per beam crossing. The interactions are spread
out over the long ($\sigma$=30 cm) interaction region.  The trigger must not
fire merely due to the presence of two nearby interactions. To insure this we
have imposed a requirement that the maximum impact parameter of a track not be
larger than 2 mm. The yield for events containing a $B^o\to\pi^+\pi^-$ decay as
a function of luminosity is shown in Fig.~\ref{lebrun_trig} (left). Here we do
not want the trigger rate to increase as a function of luminosity, even though
this means that the efficiency on this rare final state increases. Therefore,
a linear rise would be ideal.  On the
right side we show the probability to trigger on light quark background. We 
would like this to remain constant with increasing luminosity.
No increase occurs up to a luminosity of
10$^{32}$cm$^{-2}$s$^{-1}$, after which the probability for this particular
trigger condition increases mildly. However, the first level trigger rate is
clearly much lower than the 1\% we require until we exceed a luminosity of
$\approx 3\times$10$^{32}$cm$^{-2}$s$^{-1}$.

\subsection{Measurement of the CP violating asymmetry in $B^o\to\pi^+\pi^-$}

\begin{figure}[htb]
\begin{center}
\epsfig{figure=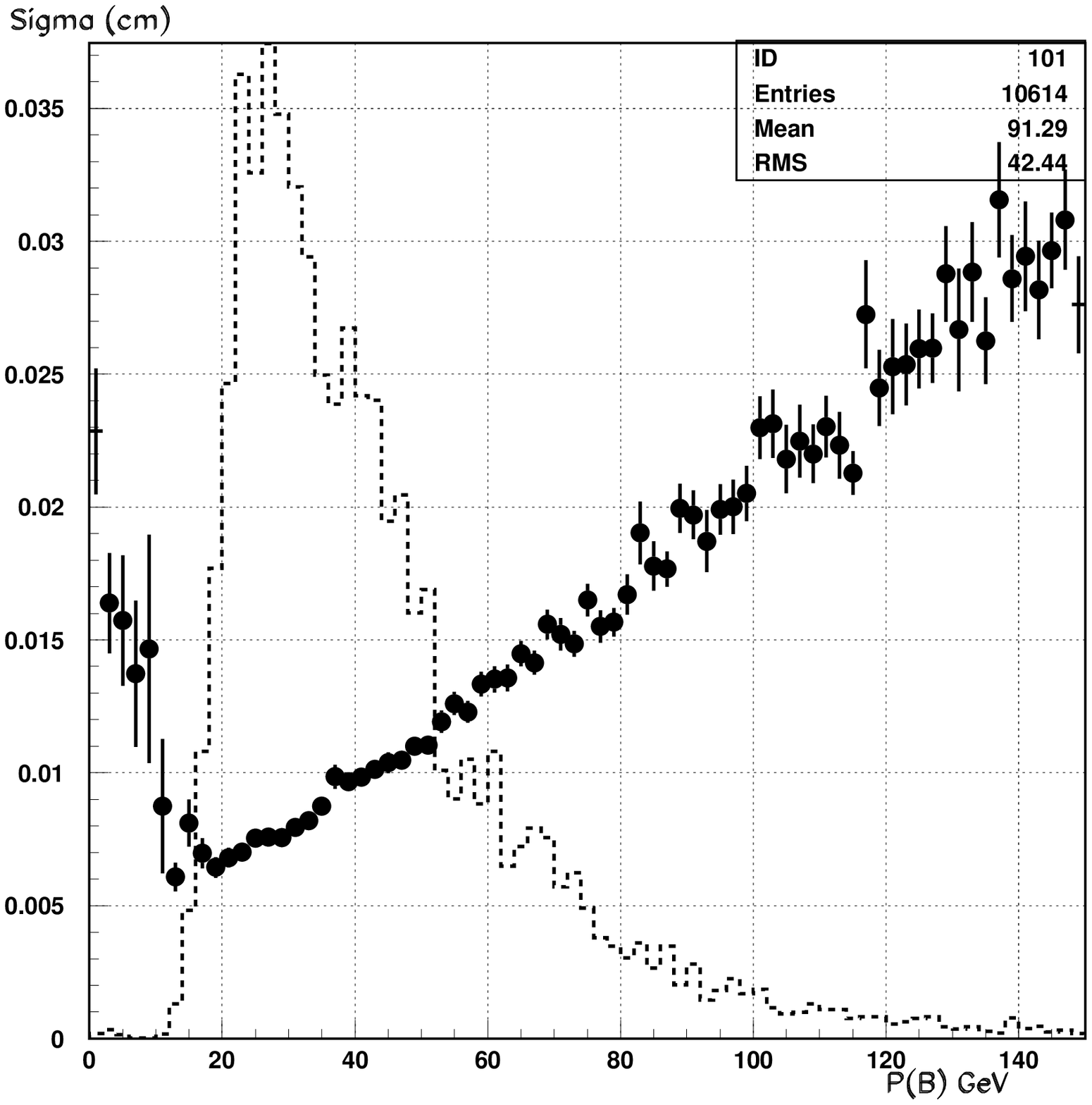,width=2.9in}
\epsfig{figure=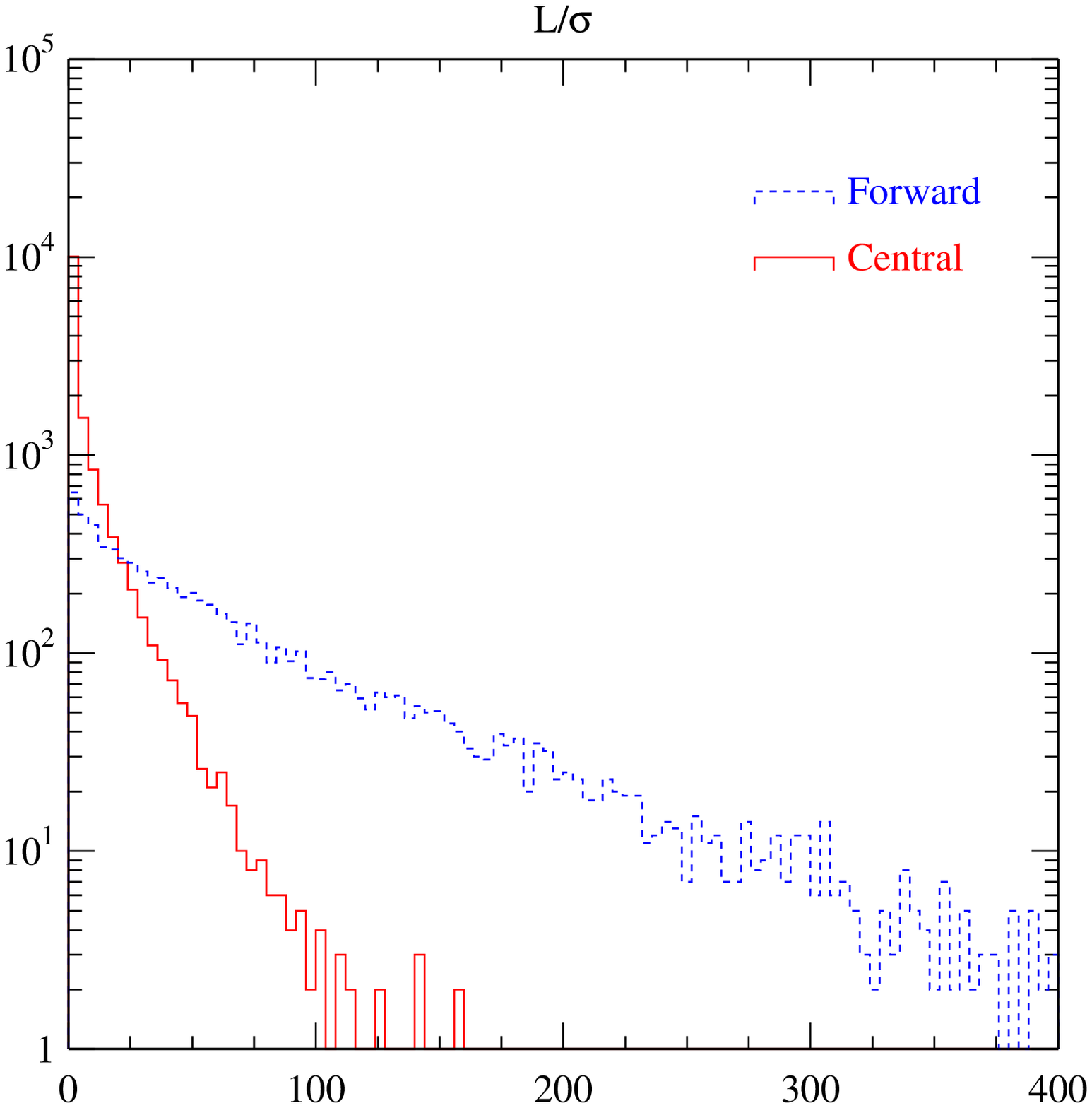,width=2.6in}
\end{center}
\caption{\label{l_over_sig}(left) The $B$ momentum distribution for events in
the detector acceptance (dashed line) and the error on the spatial distance of 
the $\pi^+\pi^-$ decay vertex from the primary vertex (solid points). (right) 
Comparison of the $L/\sigma$ distributions
for the decay $B^o\to\pi^+\pi^-$ in central and forward detectors
produced at a hadron collider with a center of mass energy of 1.8~TeV.}
\end{figure}

The trigger efficiency for this mode has already been discussed. For the
$B^o\to\pi^+\pi^-$ channel BTeV has compared the offline fully  reconstructed
decay length distributions in the forward geometry with that of a detector
configured to work in the central region. The left side plot in
Fig.~\ref{l_over_sig} shows the $B$ momentum distribution and decay distance
error as a function of $b$ momentum.

The right plot in  Fig.~\ref{l_over_sig} shows the normalized  decay
length expressed in terms of $L/\sigma$ where $L$ is the decay length and 
$\sigma$ is the error on $L$ for the $B^o\to \pi^+\pi^-$ decay \cite{procario}.

 The forward 
detector clearly has a much more favorable $L/\sigma$ distribution, which is
due to the excellent proper time resolution. The ability to keep high efficiency
in the trigger and analysis levels and devastate the backgrounds
relies primarily on having the excellent $L/\sigma$ distribution shown for the
forward detector.

\begin{figure}[htb]
\begin{center}
\epsfig{figure=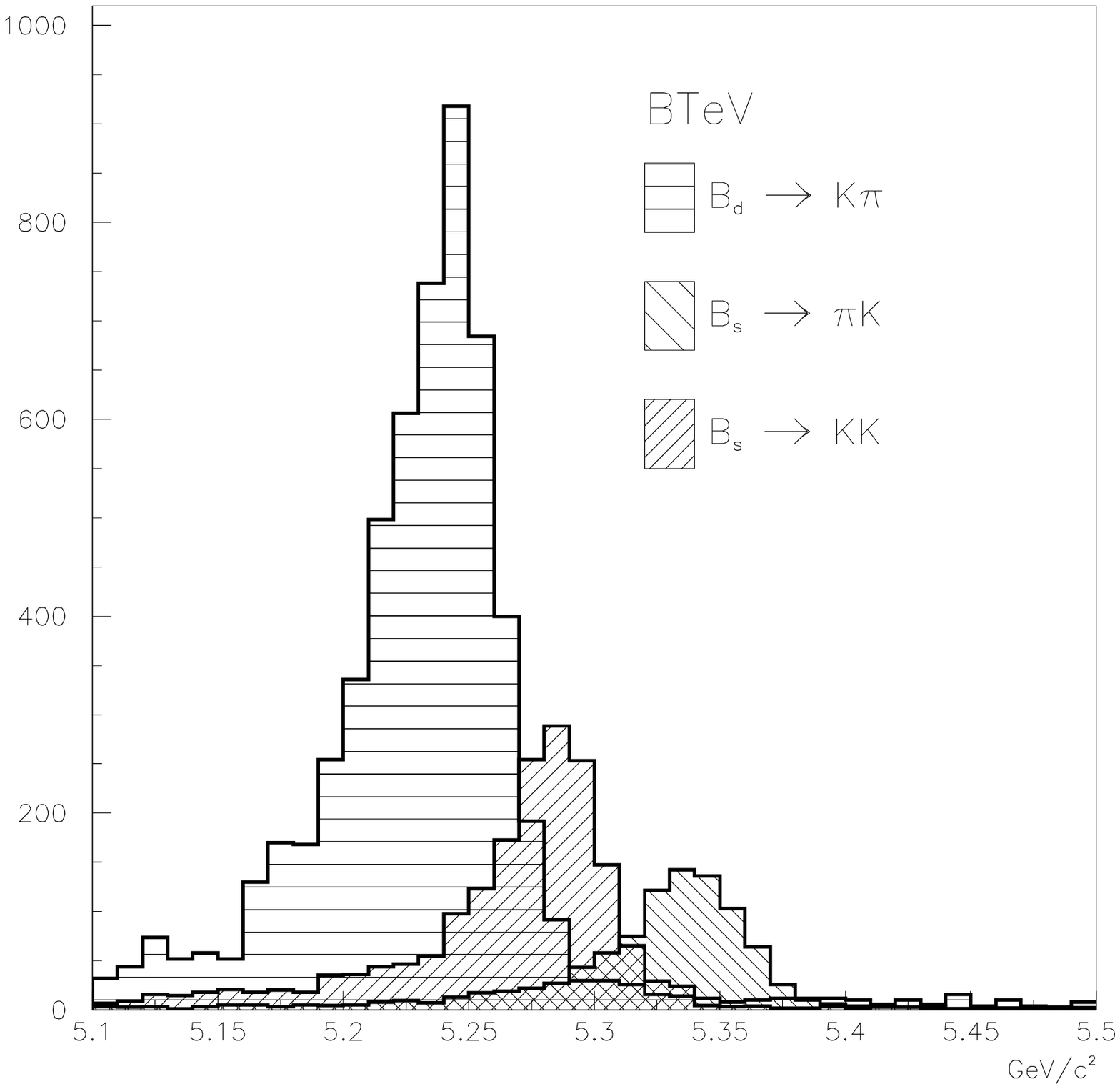,width=2.6in}
\epsfig{figure=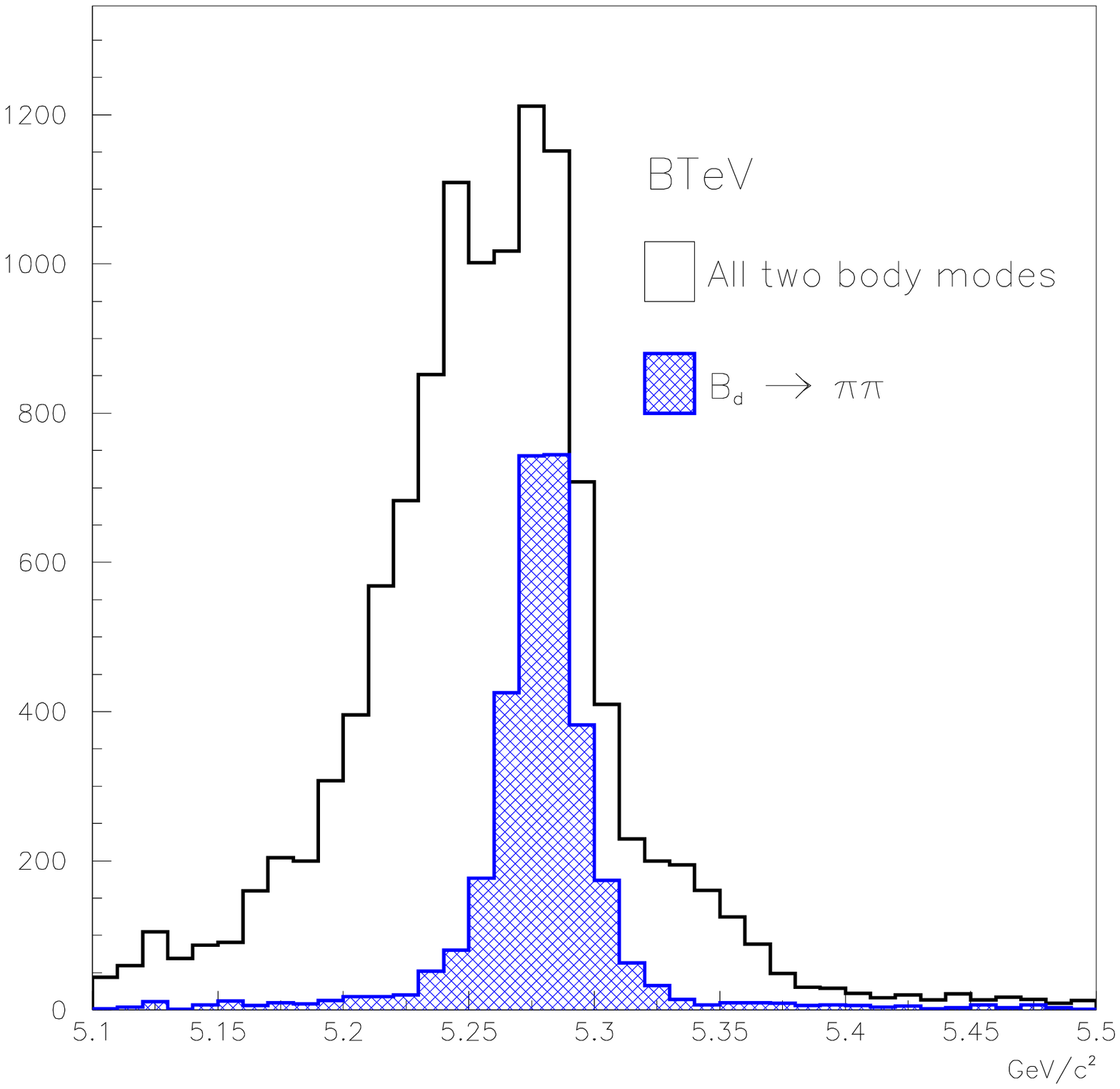,width=2.6in}
\end{center}
\caption{\label{pipi_nopid} Invariant mass distributions of all $B\to h^+h^-$
final states, where $h$ denotes either a pion or kaon, and the mass is computed
assuming that both tracks are pions. The plot on the left shows the individual
background channels and the one on the right shows the sum of all channels 
properly
normalized (see text) to the $\pi^+\pi^-$ signal.}
\vspace{1 cm}
\end{figure}

For this analysis $L/\sigma$ is required to be $>15$. Each pion track is
required to miss the primary vertex by a distance/error $>5\sigma$ and the
$B^o$ candidate is required to point back to the primary with a distance/error
$<2\sigma$. Furthermore, each track must be identified as a pion and
not a kaon in the RICH detector. Without particle identification it is
impossible to distinguish $B^o\to\pi^+\pi^-$ from the combination of $B^o\to
K^{\pm}\pi^{\mp}$, $B_s\to K^+K^-$ and $B_s\to K^{\pm}\pi^{\mp}$, as is shown
on Fig.~\ref{pipi_nopid}. Here $\cal{B}$$(B^o\to K^{\pm}\pi^{\mp})$ is taken as
$1.5\times 10^{-5}$ and $\cal{B}$$(B^o\to \pi^+\pi^-)$ is taken as $0.75\times
10^{-5}$, from recent CLEO measurements \cite{CLEOpp}. The $B_s$ decay into
$K^+K^-$ is assumed to have the same rate as the $B^o$ decay into
$K^{\pm}\pi^{\mp}$, and the $B_s$ decay into $K^{\pm}\pi^{\mp}$ is assumed to
have the same rate as the $B^o$ decay into $\pi^+\pi^-$.

Using the good particle identification, BTeV predicts that they can measure the
CP violating asymmetry in $\pi^+\pi^-$ to $\pm 0.013$ as detailed in
Table~\ref{tab:pipi}.

\begin{table}
\caption{Numbers entering into the accuracy in measuring the CP violating
asymmetry in $B^o\to \pi^+\pi^-$. \label{tab:pipi}}
\begin{center}
\begin{tabular}{lcc}
Quantity   &  Value    \\ \hline
Cross section & 100 $\mu$b      \\
Luminosity & $2\times 10^{32}$ \\
\# of $B^o$/2$\times 10^7$s, $\cal{L}$ leveled  &  $2.8\times 10^{11}$ \\
$\cal{B}$$(B^o\to \pi^+\pi^-)$  & $0.75 \times 10^{-5}$ \\
Reconstruction efficiency & 0.08 \\
Triggering efficiency (after all other cuts) & 0.72 \\
\# of $\pi^+\pi^-$ & 128,000\\
$\epsilon D^2$ for flavor tags {\small($K^{\pm}$, $\ell^{\pm}$, same + opposite
sign jet tags)}   & 0.1 \\
\# of tagged $\pi^+\pi^-$ & 12,800\\
Signal/Background & 0.9 \\
Error in asymmetry (including background) & $\pm 0.013$ \\ 
\end{tabular}

\end{center}
\end{table}

\subsection{Flavor tagging}

We have assumed a flavor tagging efficiency of 10\%. Actually our studies show
that we probably can achieve a higher efficiency. The usual definitions are:
$N$ is the number of reconstructed signal events, $N_R$ is the number of right
sign flavor tags, $N_W$ is the number of wrong sign flavor tags, $\epsilon$ is
the efficiency (given by $[N_R+N_W]/N$) and $D$ is the dilution (given by
$[N_R-N_W]/[N_R+N_W]$). The quantity of interest is $\epsilon D^2$ which when
multiplied by $N$ gives the effective number of events useful for the
calculation of an asymmetry error.

We have investigated the feasibility of tagging kaons using a gas Ring
Imaging Cherenkov Counter (RICH) in a forward geometry and compared it with
what is possible in a central geometry using Time-of-Flight counters with
good, 100 ps, resolution. For the forward detector the momentum coverage
required is between 3 and 70 GeV/c. The lower momentum value is determined by
our desire to tag charged kaons for mixing and CP violation measurements, while
the upper limit comes from distinguishing the final states $\pi^+\pi^-$,
$K^+\pi^-$ and $K^+K^-$. 
The momentum range is much lower in the central detector
but does have a long tail out to about 5 GeV/c.  Either C$_4$F$_{10}$ or
C$_5$F$_{12}$ have pion thresholds of about 2.5 GeV/c. The  kaon and proton
thresholds for the first gas are 9 and 17 GeV/c, respectively. 

\begin{figure}[htb]
\vspace{-1.3cm}
\centerline{\epsfig{figure=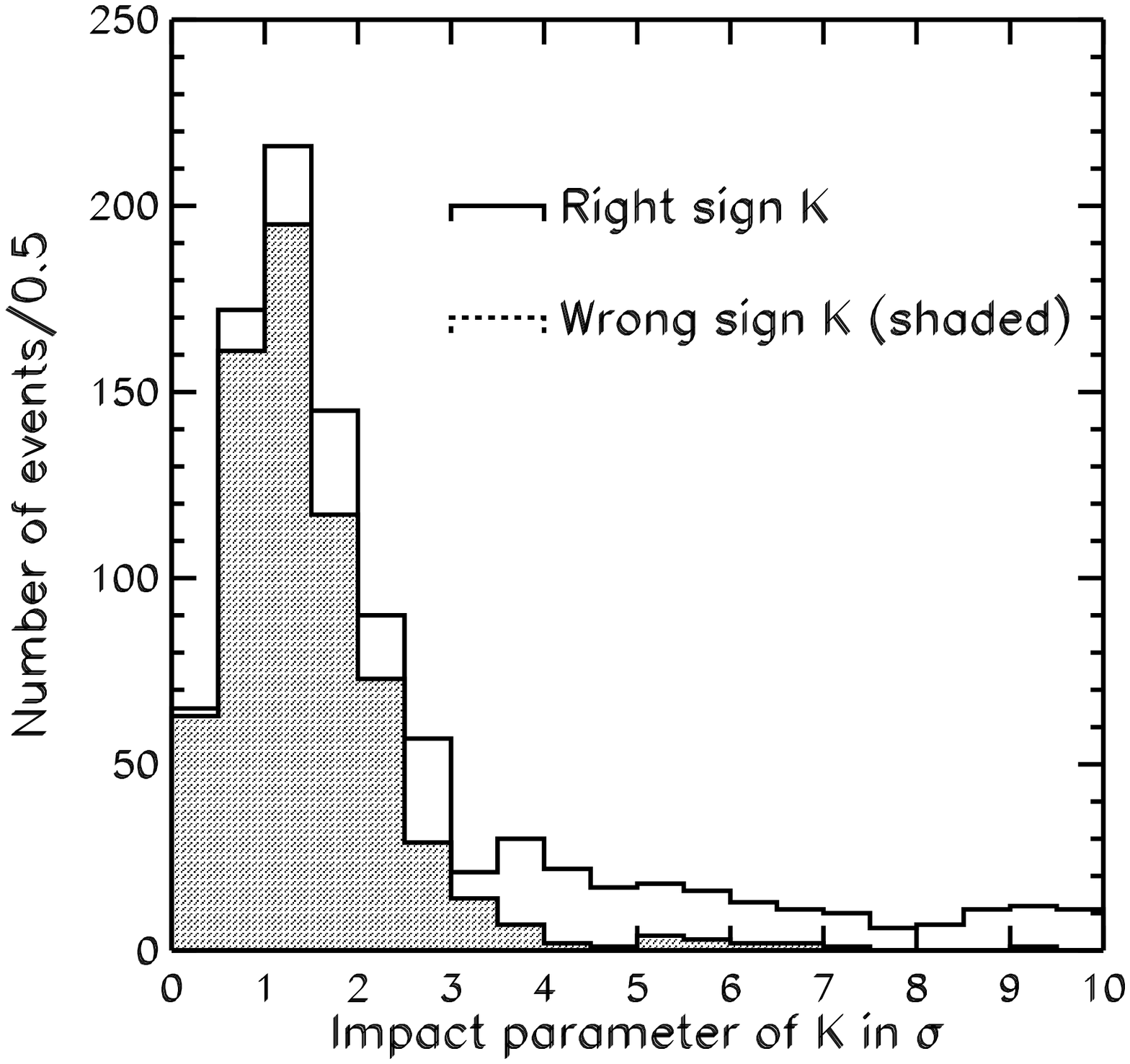,height=2.8in}
\epsfig{figure=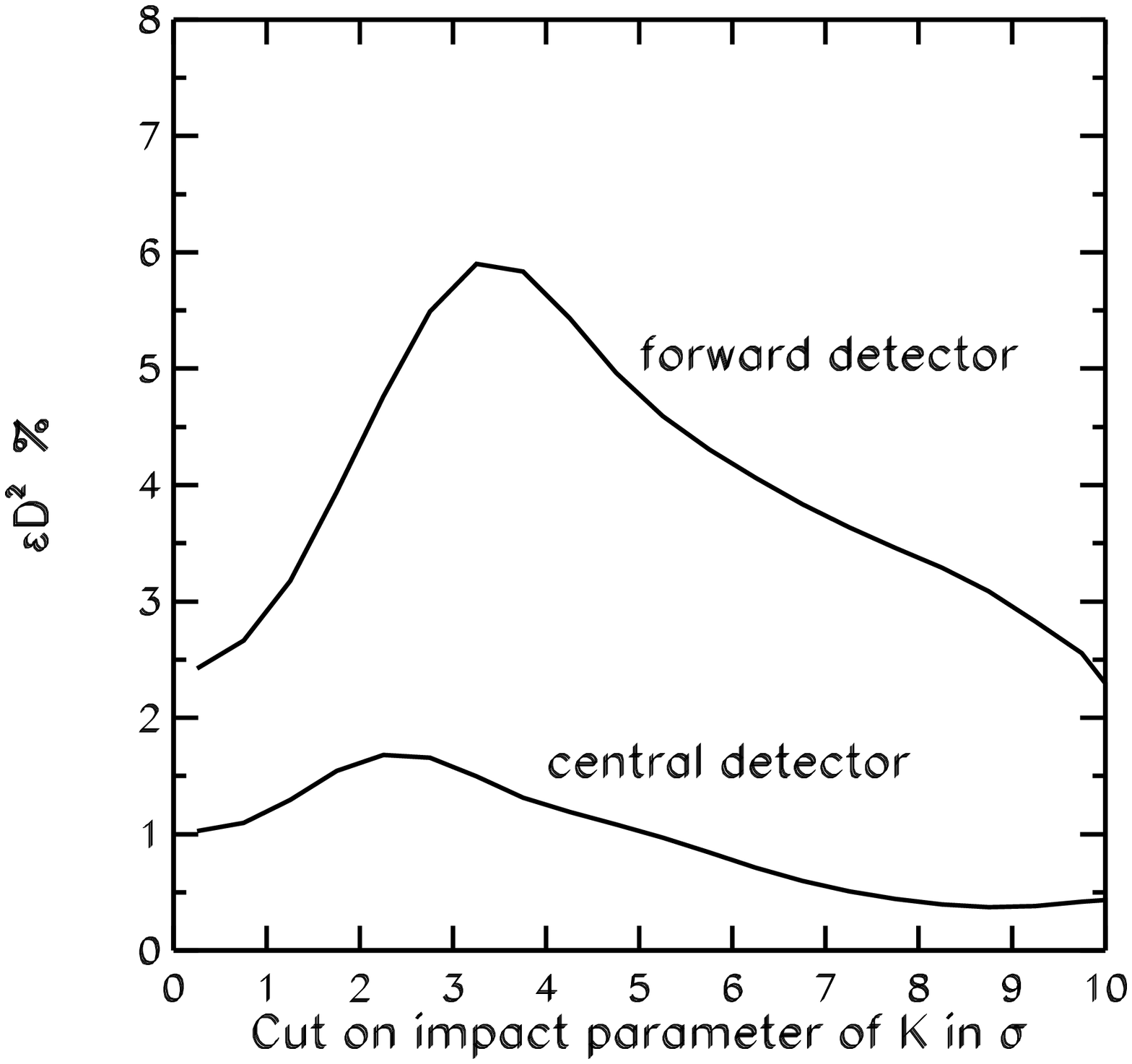,height=2.8in}}
\caption{\label{ipk}(left) $L/\sigma$ distributions in BTeV for $K^{\pm}$ impact
parameters for right sign (unshaded) and wrong sign (shaded) tags. (right) 
Overall $\epsilon$D$^2$ 
values from kaon tagging for a forward detector
containing a flourine based RICH versus a central detector with 100 ps time of
flight resolution as a function of kaon impact parameter in units of
$L/\sigma$.
(Protons and $b\overline{b}$ mixing have been ignored in both cases.)}
\end{figure}

The BTeV RICH was simulated using the current C0 geometry with MCFast.
Fig.~\ref{ipk} shows the number of identified kaons plotted versus their impact
parameter divided by the error in the impact parameter for both  right-sign and
wrong-sign kaons. A right-sign kaon is a kaon that properly tags the flavor of
the other $B$ at production. We expect some wrong-sign kaons from mixing and
charm decays. Many others just come from the primary. A cut on the 
impact-parameter standard-deviation plot at $3.5\sigma$ gives an overall
$\epsilon D^2$ of 6\%. This number is reduced to 5\% because of $b\overline{b}$
mixing \cite{mixred}. Without the aerogel preradiator to  distinguish protons
from kaons below threshold we would experience an additional reduction down to
about 4\%. These numbers are for a perfect RICH system. Putting in a fake rate
of several percent, however, does not significantly change the conclusion.

The simulation of the central detector gives much poorer numbers. In
Fig.~\ref{ipk}
$\epsilon D^2$ for both the forward and central detectors are shown as a
function of the kaon impact parameter (protons have been ignored). It is
difficult to get $\epsilon D^2$ of more than 1.5\% in the central detector.

Now let us consider other tags. We have simulated muon and electron flavor tags
in our system. Although this technique is very useful at $e^+e^-$ colliders
operating at the $\Upsilon (4S)$, it is less useful here because it is
difficult to distinguish leptons from the $b\to c\to \ell^+$ decay from the
primary leptons from the $b$ quark decay. Our estimates are given in
Table~\ref{tab:tags} along with those for a central detector.

\begin{table}
\caption{The projected flavor tagging efficiencies for a central detector
similar to CDF and BTeV in units of $\epsilon D^2$.}
\begin{center}
\label{tab:tags}
\begin{tabular}[hbt]{l|ccrccr}
 & $K^{\pm}$ & $\mu^{\pm}$ & $e^{\pm}$ & SST & Jet Charge & Sum \\
\hline
BTeV  & 5\% & 1.6\% & 1.0\% & $>$2\% & 6.5\% & $>$10\% \\
Central & 0\% & 1.0\% & 0.7\% & 2\% & 3\% & $\approx$5\% \\
\end{tabular}
\end{center}
\end{table}

The two other methods considered are ``jet charge" and ``same side" tagging
(sst). We have not yet studied sst, which is using the charge of a track
closest  in phase space to the reconstructed $B$. However, CDF has measured
$\epsilon D^2$ for it to be (1.5$\pm$0.4)\% and take 2\% as their future
projection using an improved vertex detector. We have studied jet charge, which
involves taking a weighted measure of the charge of the tagging $b$ jet.
However, we incorporate information on the detachment of the tracks to help us
define the jet. CDF extrapolates to 3\% while we expect 6.5\%.
Table~\ref{tab:tags} summarizes our projected tagging efficiencies.

\subsection{Measurement of $B_s$ mixing}

BTeV has studied the feasibility of measuring the $B_s$ mixing parameter $x_s$
= $\Delta m_s/\Gamma_s$. This measurement is key to obtaining the right side of
the unitarity triangle shown in Fig. 1. Current limits on $B_s$ mixing from LEP
give $x_s>15$ \cite{LEP-B}. Recall that for $B_d$ mesons, $x$ = 0.73. The
oscillation length for $B_s$ mixing is at least a factor of 20 shorter and may
approach a factor of 100!

BTeV has investigated two final states that can be used. The first, $\psi
K^{*o}$, $\psi\to \mu^+\mu^-$ and $K^{*o}\to K^-\pi^+$, has several advantages.
It can be selected using either a dilepton or detached vertex trigger.
Backgrounds can be reduced in the analysis by requiring consistency with the 
$\psi$ and $K^{*o}$ masses. Furthermore, it should have excellent time
resolution as there are four tracks coming directly from the $B$ decay vertex. 
The resolution in 
proper time is 42 fs.
The one disadvantage is that the decay is Cabibbo suppressed, the Cabibbo
allowed channel being $\psi\phi$ which is useless for mixing studies. The
branching ratio therefore is predicted to have the low value of $8.5\times
10^{-5}$.

\begin{figure}[htb]
\vspace{-.03cm}
\centerline{\epsfig{figure=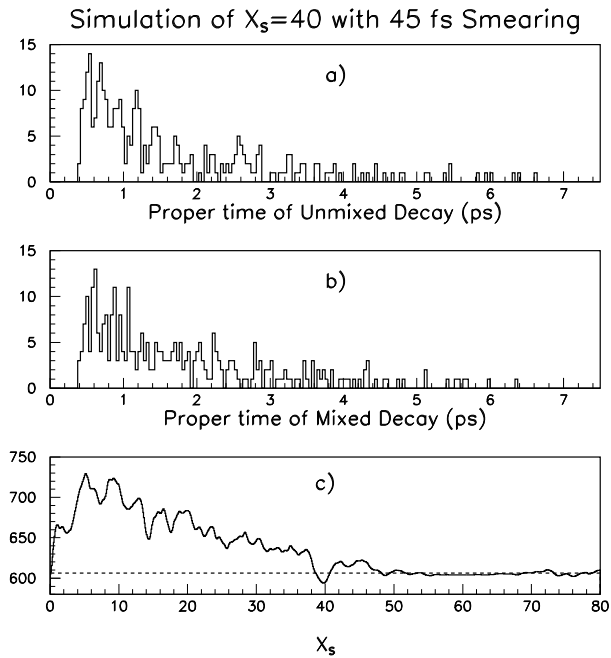,width=3.6in}}
\caption{\label{psikstime2} The observed decay time distributions for
$B_s\to\psi K^{*o}$ generated with $x_s = 40$. Unmixed decays are shown in
(a), mixed in (b). Background and mistagging have been included. In (c) the
results of a likelihood fit to the time distributions are shown. The dashed
line shows a 5$\sigma$ difference from the best solution.}
\end{figure}

The time distributions of the unmixed and mixed decays are shown in
Fig.~\ref{psikstime2}, along with a calculation of the likelihood of there
being an oscillation as determined by fits to the time distributions.
Background and wrong tags are included. 
The fitting procedure correctly finds the input value of $x_s=40$. The danger
is that a wrong solution will be found. The dashed line shows the change in
likelihood corresponding to 5 standard deviations. If our criterion is that the
next best solution be greater than $5\sigma$, then this is the best that can be
done with one year's worth of data in this mode. Once a clean solution is found,
the error on $x_s$ is quite small, being $\pm 0.15$ in this case.

\begin{figure}[htb]
\vspace{-.03cm}
\centerline{\epsfig{figure=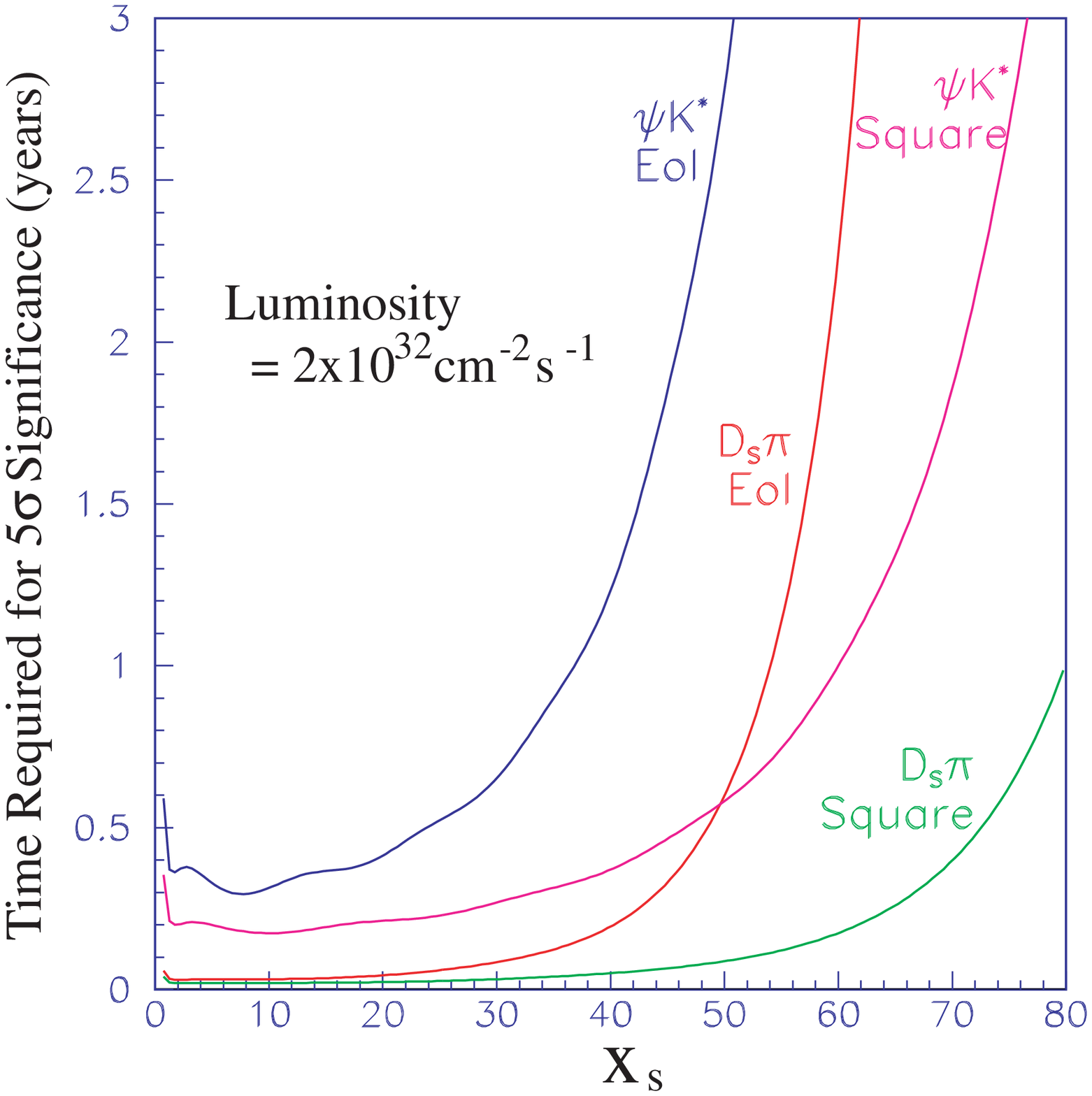,width=3.5in}}
\caption{\label{mix_summary} The $x_s$ reach for both $D_s^+\pi^-$ and
$\psi K^{*o}$ decays of the $\overline{B}_s$. The EOI detector has a 
12 mm gap between upper and lower halves of the pixel detector, while the
square hole has a 12x12 mm$^2$ hole. The calculation is for non-leveled
luminosity.}
\end{figure}

BTeV has also investigated the $D_s^+\pi^-$ decay of the $\overline{B}_s$, with
$D_s^+\to\phi\pi^+$. It turns out that the lifetime resolution is 45 fs, almost 
the
same as for the $\psi K^{*o}$ decay mode. Since the predicted branching ratio
for this mode is 0.3\%, we obtain 19200 events in one year of running, with a
signal to background of 3:1. Fig.~\ref{mix_summary} shows the $x_s$ reach
obtainable for a $5\sigma$ discrimination between the favorite solution and the
next best solution, for both decay modes. The background is assumed to be 20\%
and the flavor mistag fraction is taken as 25\%. The tagging efficiency is
taken as 10\%. The absicca gives the number of years of running, where one
year is $10^7$ seconds. The pixel system with the 12 mm gap is called the EOI 
detector here.

The other detector configuration that we simulated has the pixels configured 
around the beam leaving 
a 12 mm $\times$ 12 mm square hole. This detector has better efficiency and time 
resolution (see 
section V) and now has become the BTeV baseline.

The $x_s$ reach is excellent and extends over the entire predicted Standard
Model range. 

\subsection{ Measurement of $\gamma$ }

	The angle $\gamma$ could in principle be measured using a CP eigenstate
of $B_s$ decay that was dominated by the $b\to u$ transition. One such decay
that has been suggested is $B_s\to \rho^o K_s$. However, there are the same
``Penguin pollution" problems as in $B^o\to \pi^+\pi^-$, but they are more
difficult to resolve in the vector-pseudoscalar final state. (Note, the 
pseudoscalar-pseudoscalar final state here is $\pi^o K_s$, which does not
have a measurable decay vertex.)

	Fortunately, there are other ways of measuring $\gamma$. CP eigenstates
are not used, which introduces discrete ambiguities. However, combining
several methods should remove these.

\begin{figure}[htb]
\centerline{\epsfig{figure=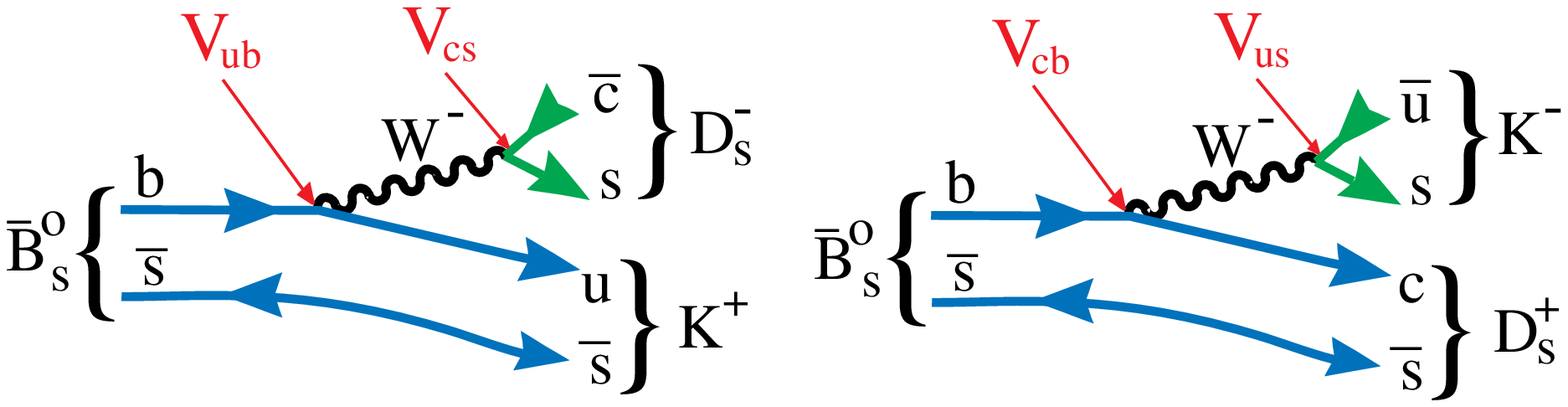,height=1.8in}}
\vspace{-.15cm}
\caption{\label{DsK} Two diagrams for $\overline{B}_s^o \to D_s^{\pm}K^{\mp}$.}
\end{figure}

  We have studied
  three methods of measuring $\gamma$.
  The first method uses the decays $B_s \rightarrow D_s^{\pm} K^{\mp}$ where a 
  time-dependent CP violation can result from
  the interference between the direct decay and the mixing induced decay 
\cite{Aleks}. Fig.~\ref{DsK} shows the two
direct decay processes for $\overline{B}^o_s$. 

   Consider the following time-dependent rates for neutral $B$ mesons to 
non-CP eigenstates via two different processes that can be separately measured 
using flavor tagging 
of the other $b$: 
\[  
  {\Gamma}(B_s\rightarrow f) = |M|^2 e^{-{t}}
  \{ \cos^2(xt/2) + \rho^2\sin^2(xt/2) - \rho{\sin(\phi+\delta)}\sin(xt) \}  
\]  
\[  \Gamma(\bar{B_s}\rightarrow \bar{f}) = |M|^2 e^{-{t}}
  \{ \cos^2(xt/2) + \rho^2\sin^2(xt/2) + \rho{\sin(\phi-\delta)}\sin(xt) \}  \]
\[  \Gamma(B_s\rightarrow \bar{f}) = |M|^2 e^{-{t}}
  \{ \rho^2\cos^2(xt/2) + \sin^2(xt/2) - \rho{\sin(\phi-\delta)}\sin(xt) \}  \]
\[  \Gamma(\bar{B_s}\rightarrow f) = |M|^2 e^{-{t}}
  \{ \rho^2\cos^2(xt/2) + \sin^2(xt/2) + \rho{\sin(\phi+\delta)}\sin(xt) \},  
\]
where  $M = \langle~f|B\rangle$, 
        $\rho = \frac{\langle~f|\bar{B}\rangle}{\langle~f|{B}\rangle}$,
        $\phi$ is the weak phase between the two amplitudes and
        $\delta$ is the strong phase between the two amplitudes.
   The three parameters $\rho$, $\sin(\phi+\delta), \sin(\phi-\delta)$
   can be extracted 
   from a time-dependent study if $\rho=O(1)$. 
        
In the case of $B_s$ decays  where $f=D_s^+ K^-$ and  $\bar{f}= D_s^- K^+$,
the weak phase is $\gamma$. The decay modes $B_s \rightarrow D_s^+ K$, $D_s^+
\rightarrow \phi \pi^+$, $\phi  \rightarrow K^+ K^-$, or $D_s\rightarrow
K^{*o}K^+$,  were simulated. For the $\phi\pi^+$ mode, the  combined geometric
acceptance and reconstruction efficiency is 5.2\% with S/B=10 \cite{Bsdsk}, and
the trigger efficiency is 67\%. In the $K^{*o} K^+$ mode the geometric and
reconstruction efficiency is 5.9\% and the trigger efficiencies and signal to
background are same as in the $\phi\pi^+$ mode. Using the branching fractions
predicted by Aleksan \cite{Aleks} and assuming a tagging efficiency $\epsilon =
15\%$ we expect  8900 events in 2$\times 10^7$ s.
    
\begin{figure}[htb]
\centerline{\epsfig{figure=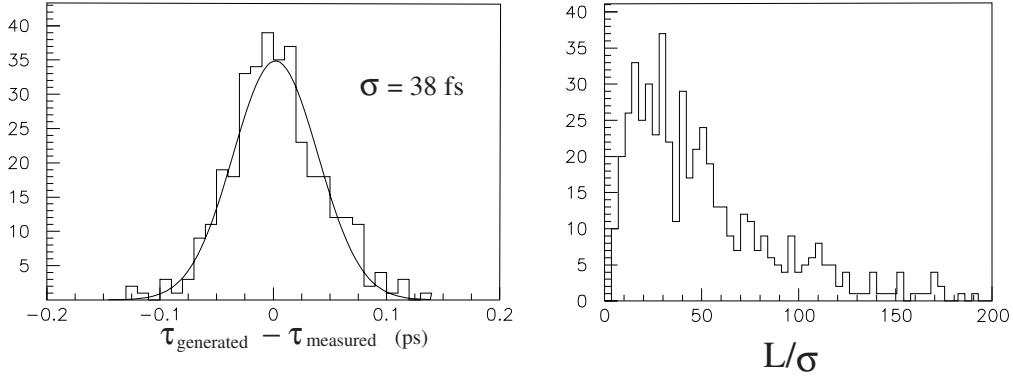,height=2in}}
\vspace{-.15cm}
\caption{\label{phipi_p} (left) The generated minus measured proper time 
distribution for $\overline{B}_s^o \to D_s^{\pm}K^{\mp}$, $D_s^+\to\phi\pi^+$.
(right) The distribution in $L/\sigma$ for this mode.}
\end{figure}

The decay time resolution and the detachment of the decay vertex from the
primary production vertex are shown in Fig.~\ref{phipi_p} for the
$D_s^+\to\phi\pi^+$ decay mode. The distributions for the $K^{*o}K^+$ mode are
similar. 

   Using the measured values of S/B and time resolution, a mistag rate of 25\%,
   and $x_s$=20,
   a mini-Monte Carlo was used to generate the extracted value of $\gamma$ for
   an ensemble of experiments each with 8900 signal events,
   for various sets of input parameters $\rho$, $\sin(\gamma+\delta), 
\sin(\gamma-\delta)$.
   A maximum likelihood fit was then used to extract fitted values of the 
parameters.
  
\begin{figure}[htb]
\centerline{\epsfig{figure=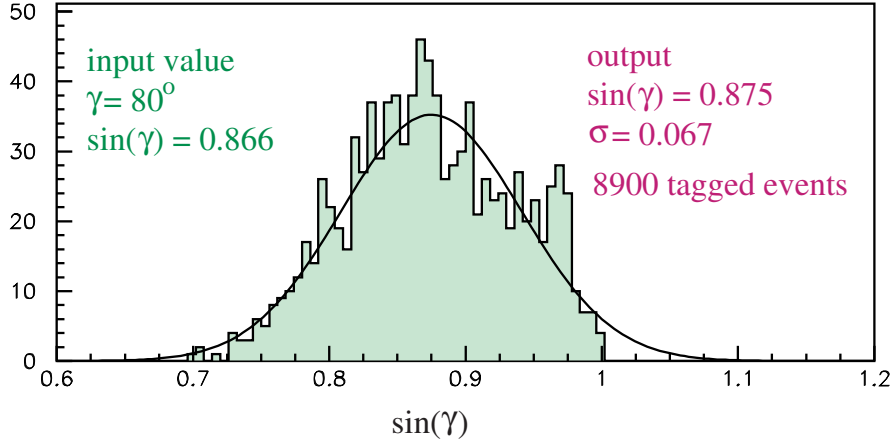,height=2.3in}}
\caption{\label{sin_g_p}  Results of determining $\gamma$ for many different
``experiments" with input value of $\gamma$=80$^{\circ}$ using the decay
time  
distributions for $\overline{B}_s^o \to D_s^{\pm}K^{\mp}$.}
\end{figure}

   Fig.~\ref{sin_g_p} shows the distributions of the parameters 
   with input values $\rho$=0.5, $\sin\gamma=0.866$ and $\cos\delta=0.7$.
   Assuming that $\sin\gamma >0$ then $\sin\gamma$ can be determined up to a 
two-fold
   ambiguity, hence $\gamma$ up to a four-fold ambiguity.
   
Another method  for extracting $\gamma$ has been proposed by Atwood, Dunietz
and Soni \cite{sad}, who refined a suggestion by Gronau and Wyler \cite{gronau}.    
A large CP asymmetry can result from the interference
of the decays $B^- \rightarrow K^- D^0, D^0 \rightarrow f$ and  $B^-
\rightarrow K^- \overline{D}^0, \overline{D}^0 \rightarrow f$, where $f$ is a
doubly Cabibbo suppressed decay of the $D^0$   (for example $f= K^+\pi^-,
K\pi\pi$, etc.)  Since $B^- \rightarrow K^- \overline{D}^0$ is color-suppressed
and $B^-  \rightarrow K^- D^0$ is color-allowed, the overall amplitudes for the
two decays are expected to be approximately equal in magnitude.  The weak phase
difference between them is $\gamma$. To observe a CP asymmetry there must also
be a non-zero strong phase between the two amplitudes. It is necessary to
measure the branching ratio  ${\cal B}(B^- \rightarrow K^- f)$ for at least 2
different states $f$ in  order to determine $\gamma$ up to discrete
ambiguities.  We have examined the decay modes $B^- \rightarrow K^- [K^+
\pi^-]$ and $B^- \rightarrow K^- [K^+ 3\pi]$. The combined geometric acceptance
and reconstruction efficiency was found to be 6.6\% for the $K \pi$ mode and
5.5\% for $K 3\pi$ with a signal to background of about 1:1.  The trigger
efficiency is approximately 70\% for both modes. The expected number of
$B^{\pm}$ events in $10^7$~s is 2400 in the $K \pi$ mode and 4200 in the $K 
3\pi$
mode. With this number of events we expect to be able to measure $\gamma$  (up
to discrete ambiguities)  with a statistical error of about  ${\pm}8^{\circ}$
in one year of running at ${\cal L}=2\times 10^{32} {\rm cm}^{-2}{\rm
s}^{-1}$.  The overall sensitivity depends on the actual values of $\gamma$ and
the strong phases.
 
The next method, described by Gronau and Rosner \cite{GR} and Fleischer and
Mannel \cite{FM}, uses $B^0 \rightarrow K^+ \pi^-$ and $B^+ \rightarrow K^0
\pi^+$ decays.  It is particularly promising as it may complement other methods
by excluding some of the region around $\gamma=\pi/2$.  We expect to
reconstruct 3600 $B^{\pm}\rightarrow K_s \pi^{\pm}$ with  S/B=0.5 and 29000
$B^0/\overline{B}^0 \rightarrow K^{\pm} \pi^{\mp}$ with S/B=3.  Gronau and
Rosner estimate a measurement of $\gamma$ to $10^{\circ}$ with 2400 events in
each channel \cite{Ros}, however there has been much  theoretical discussion
about the effects of isospin conservation and rescattering  which casts doubt
on this method \cite{Gerard}\cite{FK}\cite{MNeub}\cite{Atsoni}. There have been
several suggestions, however, on how to measure these effects \cite{KK}, and
this method may turn out to be useful. 

\section{Decay Time Resolution}

In all of these studies we have assumed that we would have 9 $\mu$m spatial 
resolution in each track hit in the pixel plane. We now address the question of
whether or not this  is reasonable.

The parameters affecting the pixel resolution include the size of the pixel,
the use of binary (one  bit) versus analog (4 bits) information, threshold and
gain variations, and the use of electrons or holes  as charge carriers, since the
drift velocity for electrons is three times that for holes. For  different
incident angles of tracks on the pixels the charge sharing is affected by the
magnetic  field. In BTeV we use a 1.5 T dipole field.

\begin{figure}[htb]
\centerline{\epsfig{figure=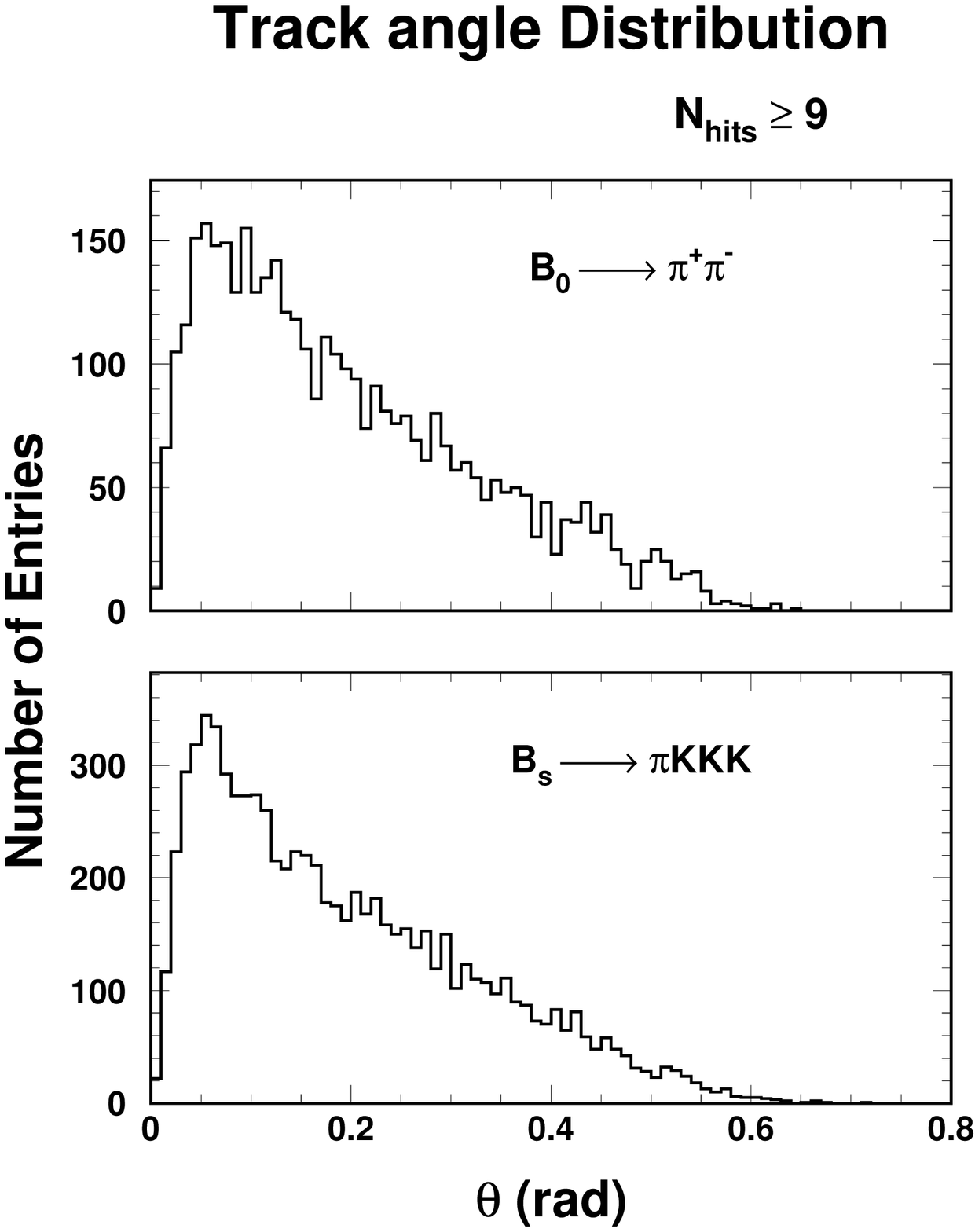,height=3.2in}
\epsfig{figure=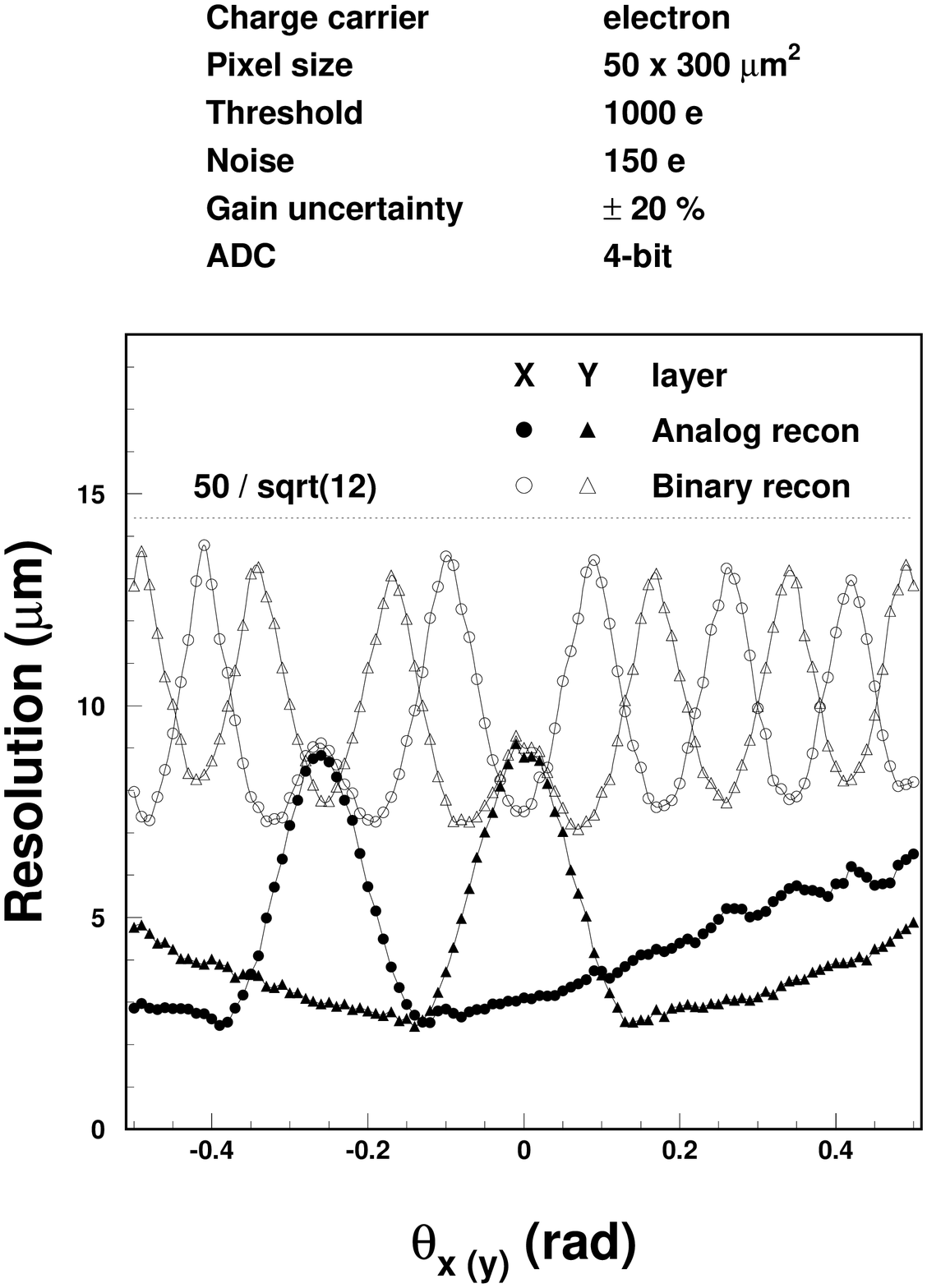,height=3.3in}}
\caption{\label{theta_w}(left) The distribution of track angles with respect
to the pixel planes for tracks with at least 9 hits for $B^o\to\pi^+\pi^-$
and $B^o\to \pi^+ K^- K^+ K^-$ decays. (right) The spatial resolution for the X
(bend plane) and Y directions, separately, as a function of projected angle,
$\Theta_{x,(y)}$, for binary and 4-bit analog readout using a pixel size
of 50$\times$300 $\mu$m$^2$. }
\end{figure}

In Fig.~\ref{theta_w}(a) we show the track angle distribution for two $B$
final  states, the two-body state  $\pi^+\pi^-$ and the four-body state $\pi^+
K^+K^-K^-$. In both cases the angular distribution  peaks at small angles,
about 50 mr and then falls slowly towards larger angles. The spatial 
resolution has been simulated as a function of angle for various pixel sizes.
The baseline size is 50  $\mu$m $\times$ 300 $\mu$m. The results of the
simulation for this size are shown in Fig.~\ref{theta_w}(b). The  magnetic
field is in the y direction in  this case. So the tracks hitting the x layers
are bent. The  resolution using binary readout is about 10 $\mu$m, while it is
about 5 $\mu$m using 4-bit analog. Note that the poorer resolution peak near 
zero degrees in the non-bend plane does exist in the bend plane, but it is
shifted toward negative incident angles. Similar results are obtained with
holes as  charge carriers.

\begin{figure}[htb]
\vspace{-1.3cm}
\centerline{\epsfig{figure=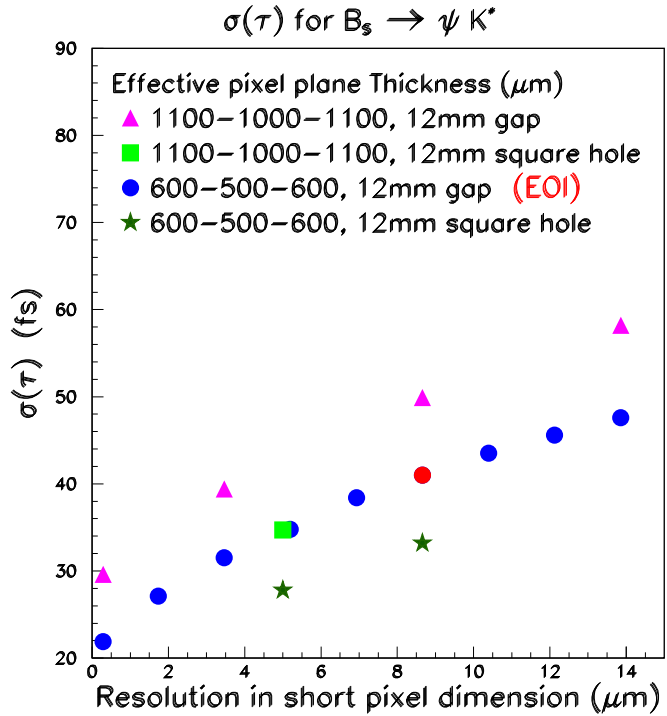,height=4in}}
\caption{\label{kut_timeres} The calculated proper time resolutions for
the decay mode $B_s\to \psi K^{*o}$, $K^{*o}\to K^-\pi^+$ as a function of 
spatial resolution in the pixel system for different detector geometries
and thicknesses. The two geometries considered are a gap of 12 mm between
two halves of the pixel system and square hole of 12mm $\times$ 12 mm. The
equivalent silicon thicknesses for the three planes are listed.}
\end{figure}

The resolution in proper time is affected by several factors. One is the
inherent pixel resolution,  as discussed above. Others include the amount of
material in the pixel system, and the distance the  pixel detector is placed
from the beam line. We take the latter as 6 mm. This distance is limited by 
the maximum amount of radiation damage we are willing to sustain. (The system
is retractable during  machine injection.) In Fig.~\ref{kut_timeres} we show
the proper time resolution achievable on the $\psi K^{*o}$ decay  of the $B_s$ 
for
several different detector geometries as a function of the spatial resolution.
The  circles represent a geometry with a 12 mm gap and equivalent silicon
thickness of 600, 500, and 600  $\mu$m for the three layers. These include the
300 $\mu$m of silicon for each layer, a radio-frequency  shield of 100 $\mu$m 
of
Al and material for electronics and cooling.

The simulations presented here have assumed a 9 $\mu$m  spatial resolution,
even though we believe that 5 $\mu$m is possible. The equivalent silicon
thickness of a three plane station is taken at 1700 $\mu$m. A detector of twice
the material thickness and 5 $\mu$m resolution would have the same time
resolution as the one we have been using. This points out the need to minimize
the material, a well known lesson.

\section{Comparisons with other experiments}

\subsection{Comparisons with $e^+e^-$ $B$-factories}

Most of what is known about $b$ decays has been learned at $e^+e^-$ machines
\cite{Bdecays}. Machines operating at the $\Upsilon (4S)$ found the first fully
reconstructed  $B$ mesons (CLEO), $B^o$-$\overline{B}^o$ mixing (ARGUS), the
first signal for the $b\to u$ transition (CLEO), and Penguin decays (CLEO).
Lifetimes of $b$ hadrons were first measured by experiments at PEP, slightly later at
PETRA, and extended and improved by LEP \cite{Bdecays}. 

The success of the $\Upsilon (4S)$ machines has led to the construction at KEK
and SLAC of two new $\Upsilon (4S)$ machines with luminosity goals in excess of
$3\times 10^{33}$cm$^{-2}$s$^{-1}$. These machines will have asymmetric beam
energies so they can measure time dependent CP violation. They will join an
upgraded CESR machine at Cornell with symmetric beam energies. These machines 
will investigate only 
$B^o$ and
$B^{\pm}$ decays, they will not investigate $B_s$, $B_c$ or $\Lambda_b$ decays.

Table~\ref{tab:pipiee} shows a comparison between BTeV and an asymmetric
$e^+e^-$ machine for measuring the CP violating asymmetry in the decay mode
$B^o\to\pi^+\pi^-$. It is clear that the large hadronic $b$ production
cross section can overwhelm the much smaller $e^+e^-$ rate.

\begin{table}
\caption{Number of tagged $B^o\to\pi^+\pi^-$ ($\cal{B}$=$0.75\times 10^{-5}$)}
\begin{center}
\label{tab:pipiee}
\begin{tabular}[t]{lrccrlr}
&$\cal{L}$(cm$^{-2}$s$^{-1})$&$\sigma$&\# $B^o /10^7$s&efficiency&
$\epsilon D^2$& \# tagged \\
\hline
$e^+e^-$&$3\times 10^{33}$&1 nb &$3.0\times 10^{7}$&0.4 &0.4 &46\\
BTeV$^{\dag}$&$2\times 10^{32}$&100$\mu$b &$1.1\times 10^{11}$&0.06 &0.1
&6400\\
BTeV$^{\ddag}$&$2\times 10^{32}$&100$\mu$b &$2.1\times 10^{11}$&0.06 &0.1
&12800\\\hline
\multicolumn{7}{l}{\dag This is for gap detector, expect increase with square
hole} \\
\multicolumn{7}{l}{\ddag Luminosity leveled, use $2\times 10^7$ s/year (gap
detector)} \\
\end{tabular}
\end{center}
\end{table}

\subsection{Comparisons with Tevatron Central Detectors}

Both CDF and D0 have measured the $b$ production cross section \cite{bcross}
and CDF has contributed to our knowledge of $b$ decay mostly by its
measurements of the lifetime of $b$-flavored hadrons \cite{CDFlife}, which are
competitive with those of LEP \cite{LEPlife} and recently through its discovery 
of the $B_c$
meson \cite{bc}. These detectors were designed for physics discoveries at large
transverse momentum. It is remarkable that they have been able to accomplish so
much in $b$ physics.

However, these detectors are very far from optimal for $b$ physics. BTeV has
been designed with $b$ physics as its primary goal. To have an efficient
trigger based on separation of $b$ decays from the primary, BTeV uses the large
$\eta$ region where the $b$'s are boosted. The detached vertex trigger allows
collection of interesting purely hadronic final states such as $\pi^+\pi^-$,
$D_s^+\pi^-$ and $D_s^+K^-$. It also allows us to collect enough charm to
investigate mixing and CP violation.

The use of the forward geometry also allows excellent charged hadron
identification with a gaseous RICH detector. This is crucial for many physics
issues such as separating $K\pi$ from $\pi\pi$, $D_s\pi$ from $D_s K$,
kaon flavor tagging etc...

\subsection{Comparison with LHC-B}

LHC-B is an experiment proposed for the LHC with almost the same physics goals
as BTeV \cite{LHC-B}. LHC-B has two advantages: the $b$ cross section is five
times higher than at the Tevatron while the total cross section is only 1.6
times as large, and the mean number of interactions per crossing is three times
lower, because the LHC has bunches every 25 ns, while the Tevatron bunches come
every 132 ns.

There are, however, many advantages which accrue to BTeV. Let us first consider
the machine specific ones. The 132 ns bunch spacing at the Tevatron makes first
level detached vertex triggering easier. It is difficult for the vertex
detector electronics in LHC-B to settle in 25 ns. The seven times larger energy
at the LHC results in a larger track multiplicity per collision which causes
trigger and tracking problems and a larger range of track momenta that
need to be analyzed. The interaction region at the LHC is relatively short,
$\sigma$=5 cm, compared with the 30 cm long region at Fermilab. This somewhat
compensates for the larger number of interactions per crossing, since the
interactions are well separated.

There are detector specific advantages for BTeV as well. BTeV is a two-arm
spectrometer, resulting a factor of two advantage. BTeV has the vertex detector
in the magnetic field which allows the rejection of high multiple scattering
(low momentum) tracks in the trigger. Furthermore, BTeV is designed around a
pixel vertex detector while LHC-B has a silicon strip detector. BTeV can put
the detector closer to the beam (6 mm versus 1 cm), and has a much more robust
tracking system which can trigger on detached verticies in the first trigger
level, while LHC-B triggers on tracks of moderate transverse momentum in their
first trigger level. 

We feel that we have more than compensated for LHC-B's initial advantages.

\section{Conclusions}

Hadron colliders have large $b$ and $c$ cross sections allowing the
opportunity for  precision measurements of CP violation and $B_s$
mixing. In our view this requires high density tracking and triggering
information that  can be provided by a state of the art pixel system.   BTeV
has been designed to fit in the new C0 interaction region at the Tevatron and 
incorporates a  pixel vertex detector,  downstream tracking, charged paricle
identification, lepton identification and photon detection. The vertex detector
enables  Level I vertex triggering and excellent time resolution on heavy
hadron decays \cite{web}.

A summary of the physics reach is shown in Table~\ref{tab:phys}. Those
simulations that have  been upgraded by using the square hole detector are so
indicated.

\begin{table} [hbt]
\caption{BTeV Physics Reach}
\begin{center}
\label{tab:phys}
\begin{tabular}[t]{cc}

Measurement & Accuracy in $10^7$ s \@ \\
            &  ${\cal L}=2\times 10^{32}$, 
$\cal{L}$ leveled \\
\hline
$x_s$ (square hole) & up to 80 \& beyond \\

$A_{CP}(B^0\rightarrow\pi^+\pi^-)$ (gap)   & $\pm 0.013$ \\

$\gamma$ using $D_s K^-$ (square hole) & $\pm 8^{\circ\dag}$ \\

$\gamma$ using $D^0 K^-$  (square hole)    & $\pm 8^{\circ\ddag}$ \\

${\cal B}(B^- \rightarrow K^-\mu^+\mu^-)$ (gap)  & 
$4\sigma$ at ${\cal B}$ of $5.4\times10^{-8}$ \\
$\sin (2\beta)$ using $B^o\to\psi K_s$ (square hole) & $\pm 0.013$\\
\hline
\multicolumn{2}{l}{\dag Assumes $\rho$=0.7, $\cos\delta$=0.7, 
$\sin\gamma$=0.5, $x_s$=20} \\
\multicolumn{2}{l}{\ddag For most values of strong phases and $\gamma$} \\
\end{tabular}
\end{center}
\end{table}

BTeV is an officially recognized R\&D project at Fermilab. Development has 
started on the pixel, trigger, RICH, muon, forward tracking and electromagnetic 
calorimeter systems. More information on BTeV can be found on the world-wide-web
\cite{web}.



\begin{thebibliography}{99}
\bibitem{mixred}
About 20\% of the $B_d$ mesons mix and 50\% of the $B_s$. However, the final
state in most $B_s$ decays contains an $s$ quark from the $b\to c\to s$ decay
chain and an $\overline{s}$ spectator quark, negating any effects from mixing.
In fact, the $B_s$ does not usefully contribute to kaon tagging in the first
place. Therefore we are left with a dilution of 8\% from mixing in kaon
tagging, taking the $B_d$ fraction as 40\%.
\end{thebibliography}
\end{document}